\documentclass{aa}
\usepackage{amsmath}
\usepackage{amsfonts}
\usepackage{booktabs}
\usepackage{subfig}
\usepackage{graphicx}
\usepackage{float}
\usepackage{subfig}
\usepackage{booktabs}
\usepackage{ulem}
\usepackage{float}
\usepackage{caption}
 
\usepackage{amsfonts}
\usepackage{amssymb}   
\usepackage{thmtools}
\usepackage{indentfirst} 
\usepackage{url}
\usepackage[colorlinks=true,linkcolor=blue,citecolor=blue]{hyperref}
\usepackage[dvipsnames]{xcolor}
\usepackage{titlesec, color}
\usepackage{ifthen}
\usepackage{etoolbox}
\usepackage{txfonts}

\usepackage{graphicx,lipsum}

\def\Msol{{\rm M}_\odot}
\def\Msolperyr{{\rm M}_\odot\,{\rm yr}^{-1}}

\def\SFRradio{\textrm{SFR}_{\textrm{radio}}}
\def\SFRsed{\textrm{SFR}_{\textrm{SED}}}
\def\gradient{\nabla \textrm{SFR}_{300}}
\def \cigale{{\texttt{CIGALE}}}

\begin{document}

\title{Probing the Timescale of the 1.4\,GHz Radio emissions as a Star formation tracer} 

\author{R. C. Arango-Toro\inst{1}\fnmsep\thanks{E-mail: \url{rafael.arango-toro@lam.fr}}, 
L.~Ciesla\inst{1}, 
O.~Ilbert\inst{1},
B.~Magnelli\inst{2},
E.~F.~Jim\'enez-Andrade\inst{3},
and
V.~Buat\inst{1}. 
}

\institute{ 
Aix Marseille Univ, CNRS, CNES, LAM, Marseille, France
\and
Universit{\'e} Paris-Saclay, Universit{\'e} Paris Cit{\'e}, CEA, CNRS, AIM, 91191, Gif-sur-Yvette, France
\and
Instituto de Radioastronom\'{ı}a y Astrof\'{ı}sica, UNAM Campus Morelia, Apartado postal 3-72, 58090 Morelia, Michoac\'{a}n, M\'{e}xico
}

  \date{Received: January 4, 2023; Accepted: 14 April, 2023}

  \abstract
  {Radio used as a star formation rate (SFR) tracer presents enormous advantages by being unaffected by dust and radio sources being pinpointed at the sub-arc-second level. But the interpretation of the low frequency 1.4\,GHz luminosity is hampered by the difficulty in modeling the cosmic ray paths in the interstellar medium, and their interactions with the magnetic field.}
  {In this work, we compare the SFR derived from radio observations, and the ones derived from spectral energy distribution (SED) modeling. We aim at better understand the behavior of the SFR radio tracer, with a specific emphasis on the link with star-formation histories. }
  {The analysis is based on a sub-sample of 1584 star-forming galaxies extracted from the COSMOS VLA 3\,GHz survey project. We used the SED modeling code Code Investigating GALaxy Emission, \cigale , with a non-parametric star formation history model (SFH) and fit the data over the wavelength range from the ultraviolet (UV) up to the mid-infrared (mid-IR). We interpret the difference between radio and SED-based SFR tracers in the light of recent gradients in the derived SFH. To validate the robustness of the results, we checked for any remaining active galaxy nuclei (AGN) contribution and tested the impact of our SFH modeling approach.}
  {Approximately 27\% our galaxies present a radio SFR ($\SFRradio$) at least ten times larger than the instantaneous SFR from SED-fitting ($\SFRsed$). This trend affects primarily the galaxies that show a declining SFH activity over the last 300\,Myr. Both SFR indicators converge toward a consistent value, when the SFHs are averaged over a period larger than 150 Myr to derive SFR$_{\textrm{SED}}$.}
  {Although the radio at low frequency 1.4\,GHz is a good tracer of the star formation activity of galaxies with constant or increasing SFH, our results indicate that this is not the case for galaxies that are quenching. Our analysis suggests that the star formation time sensitivity of the radio low frequency could be longer than 150\,Myr. Interestingly, the discrepancy between the $\SFRradio{}$ and $\SFRsed{}$ could be used as diagnostic to select post-starburst galaxies.}


   \authorrunning{Arango-Toro et al.}
   \titlerunning{Radio star formation rate timescale.}

   \maketitle
%

\section{Introduction}

Galaxies' star formation activity is closely linked to their gas reservoir, feedback processes, and metal production \citep[e.g.][]{Lilly13}. The star formation rate (SFR) is an essential probe of the instantaneous state of a galaxy.This observable quantity can be estimated from a wide range of wavelengths \citep[e.g.][]{Kennicutt98,kennicutt12}.

The various SFR diagnostics are systematically tested and compared to establish their accuracy \citep{DominguezSanchez12,Figueira22}. 
Most of the SFR diagnostics are related directly or indirectly to the emissivity of the most massive short-lived O and B stars which dominates the UV energy density. The UV SFR tracer presents the advantage of being observable from the local Universe \citep[e.g.][]{Boissier07} to the most distant galaxies \citep[e.g.][]{Finkelstein22}, and for millions of galaxies \citep[e.g.][]{Moutard20}. This tracer is sensitive to short timescale variations ($<$100Myr) of the star-formation although this is dependant on the assumed star formation history \citep[][and reference therein]{Boquien19}. But dust attenuation can not be ignored. Indeed, unobscured galaxies account for less than 20\% of the SFR density at $1<z<2$ \citep[e.g.][]{LeFloch05,Magnelli13}. 
The UV light being reprocessed by dust and emitted in infrared (IR, from 8\,$\mu$m to 1000\,$\mu$m), combining both wavelengths is acknowledged as one of the most accurate diagnostic of the SFR assuming a constant SFH over 100\,Myr \citep[][]{Bell05, Buat05,kennicutt12,Arnouts13,Buat19}. But even with a 3.5-meter class telescope as \textit{Herschel}, the sensitivity is limited to the most active star-forming galaxies (SFGs) at $z>3$. Moreover, the beam size ranging from 6.8\arcsec{} to 36.3\arcsec{} \citep[][]{Oliver12} is a limitation to pinpointing the associated optical counterpart and could bias the observed flux because of blending.

The radio wavelength can be seen as a promising alternative to estimate the SFR. The main advantage is that the radio emission is not affected by dust \citep{Condon92}. Moreover, radio detection can be located with a positional accuracy at a sub-arc-sec scale, which makes the identification of the optical counterparts more reliable than in the IR. The small size of the beam allows an efficient use of stacking techniques \citep{karim11,leslie20}. The tremendous efforts in the development of radio telescopes (e.g. Square Kilometer Array) is promising for the future use of the radio as SFR tracer. 

At high-frequency $>10-20$\,GHz, the thermal emission (free-free) in HII region is well understood to explain the radio emissivity. At this frequency range, the radio traces the star-formation over short timescales ($<10$\,Myr). However, the link between star-formation and radio emission at lower frequency (around $1$\,GHz) is more complex and less understood \citep{Condon92}. The non-thermal emission is created by highly accelerated electrons trapped in the magnetic fields in the interstellar medium (ISM), which generates synchrotron emission. The electrons are accelerated during the supernovae phase and travel the ISM \citep{Tabatabaei17}. The emission starts only when the star explodes, creating a time lag between the onset of star-formation and the radio emission. The radio emission could occur over long timescales depending on the cosmic rays lifetime and the strength of the magnetic fields. Indeed, cosmic rays are subject to several cooling processes as they propagate throughout the ISM, which are mainly caused by inverse Compton, bremsstrahlung, and ionisation losses \citep[e.g.,][]{Murphy09}. The difficulty to understand and modelize such radio emission makes its use more uncertain. Therefore, a standard approach is the use of the IR-radio correlation to estimate the SFR \citep[e.g.][]{Helou85,Magnelli15,Delhaize17,Delvecchio21}. In addition, the Active Galactic Nuclei (AGN) contribution in the radio wavelength domain could bias the result if misinterpreted as linked to star-formation. 
In the local Universe  (z $\lesssim$ 0.3) radio AGN are preferentially hosted by red and massive galaxies (M$_* \gtrsim 10^{11}$M$_{\odot}$) which could be identified and discarded \citep{Smolcic08}. But this criterion is not sufficient to remove High Excitation Radio Galaxies (HERGs) \citep{Best12,Janssen12,Gurkan15} which could be hosted by galaxies with a moderate star-formation activity. The difficulty is even more severe at higher redshift \citep{Williams15,Pracy16} with the fraction of HERGs increasing in intermediate mass galaxies. Therefore, several diagnostics must be implemented to discriminate radio AGN sources from star-forming galaxies \citep[][and references therein]{Padovani16,Magliocchetti22}.

In this paper, we will focus on he spectral energy distribution (SED) modeling techniques to better characterize the radio wavelength range as a star-formation tracer. The last decade has seen tremendous progresses in the SED-fitting techniques, allowing to extract physical parameters from a wide wavelength baseline. The first generation of SED-fitting codes were limited to the exploitation of the stellar light using multi-lambda images in visible and near-infrared (NIR), as \texttt{Hyperz} \citep{bolzonella00} or \texttt{Le Phare} \citep{Arnouts02,Ilbert06} for photometric redshifts estimation or codes focusing on the physical parameters determination \citep[][and reference therein]{Walcher11}. New generation of codes extends now to other multi-lambda domains. IR is a crucial wavelength domain to assess the contribution from dust, as included in, for instance, \cigale\ \citep{Noll09,Boquien19} or \texttt{MAGPHYS} \citep{dacunha08}. Information extracted from narrow spectral features can bring information on the physical state of HII regions, as in \texttt{BEAGLE} \citep{Chevallard16}. A recent progress has been the ability to model more complex star-formation histories (hereafter SFH). While first limited to simple analytical representation of the SFH, studies have added the possibility of a recent change as a quenching or a burst event \citep[][]{Ciesla17,Schreiber18,Aufort20,Ciesla21}. Even more powerful techniques have been developed to include varying SFH, with non-parametric modeling, as in \texttt{PROSPECTOR} \citep{leja17,Tacchella22} or \cigale\ \citep{Ciesla22}. In this paper, we compare the SFR derived from radio emission and SED modeling techniques using the multi-wavelength COSMOS data. We aim at better understanding the sensibility of the radio taken as SFR tracer, depending on the timescale considered. Our work relies on the SFH estimated from a SED modeling technique, assuming non-parametric modeling of the SFH.
The article is organized as follows. Sect.~\ref{data_sample} describes the data set employed in this work which consists of a seminal galaxy sample from the work of \citet{Jimenez19}. This sample includes several physical parameters, such as the radio SFR indicator obtained through the infrared-radio correlation. In Sect.~\ref{s-m}, we present the Code Investigating GALaxy Emission, \cigale\,  which has been employed to recover the star-forming properties from broad and narrow-band SED modelling, this in a wavelength range coming from the near ultra-violet (NUV), up to the far infra-red (FIR). Sect.~\ref{res} illustrates the results from the data analysis to eventually discuss and conclude in the last section.
We adopt the standard $\Lambda$CDM cosmology with $\Omega_\textrm{m}=0.3$, $\Omega_\Lambda=0.7$ and $H_0=70$~km s$^{-1}$ Mpc$^{-1}$. We use the initial mass function (IMF) from \cite{Chabrier03}. Magnitudes are given the AB system \citep{oke_absolute_1974}.
\section{Galaxy sample} \label{data_sample}
\subsection{VLA COSMOS 3 GHz project}
The VLA COSMOS 3\,GHz survey \citep{Smolcic17} counts for 348\,h of observations using the \textrm{Karl G. Jansky Very Large Array} with a total of 192 individual pointing performed to achieve an uniform rms over the two square degrees of the COSMOS field where each pointing was imaged individually using a circular restored beam with a Full Width at Half Maximum (FWHM) of 0.75\,arcsec. The final mosaic was produced using a noise-weighted mean of all the individually imaged pointing, reaching a median rms deviation of 2.3\,$\mu$Jy\,beam$^{-1}$. The final catalog presents a total of 10,830 radio sources down to 5$\sigma$.

\subsection{ \citet{Jimenez19} sample } \label{sample} 

We use a VLA COSMOS 3\,GHz sub-sample of star-forming galaxies drawn by \citet{Jimenez19} where they first ran \texttt{PyBDSF} \citep{PyBDSF07} to extract the radio sources from the mosaic to create a catalog with flux and size measurements. Then, this preliminary catalog was cross-matched with the original one from \citet{Smolcic17}, to select sources that are detected in both catalogs  (this helping to reduce the spurious fraction on the final sample), where the number of common sources lead to a initial catalog with 9,223 galaxies.

\subsubsection{Star-forming galaxies selection} \label{AGN-selec}
To exclude AGN from their initial sample, \citet{Jimenez19} cross-matched the initial catalog of 9,223 sources with the radio source population catalog from \citet{smolcic17b}, who presented a sample of ``pure'' star-forming galaxies via the following selection criteria to identify AGN:

\begin{itemize}
    \item the intrinsic [0.5-8]\,keV X-ray luminosity is greater than $L_X=10^{42}$\,erg s$^{-1}$ \citep[e.g.,][]{Szokoly04};
    \item the flux throughout the four IRAC bands (3.6, 4.5, 5.8, and 8\,$\mu$m) displays a monotonic rise and follows the criterion proposed by \citet{Donley12};
    \item an AGN component significantly improves the fitting of their optical to millimeter SED \citep[see for instance][]{dacunha08,Berta13,Delvecchio14};
    \item $M_{NUV}-M_R$, i.e., rest-frame NUV minus $r+$ band, is greater than 3.5, which select quiescent galaxies from the red sequence \citep{Ilbert10};
    \item the observed radio emission L$_{1.4\,{\rm GHz}}$ exceeds that expected from the host galaxy SFR$_{IR}$ \citep[estimated via IR SED fitting,][]{Delvecchio17}.
\end{itemize}

Then \citet{Jimenez19} applied a cut in redshift by considering a distribution between $0.35<z<2.25$. They excluded sources whose radio emission is modelled by more that one Gaussian component, this to avoid complex radio morphology, leading to a reduced sample of 3,184 radio sources. Finally two more selection criteria were applied by keeping galaxies with distances to main sequence $\Delta{MS} > -0.3$, and galaxies with stellar mass above the mass limit defined for each stellar mass bin $(\textrm{log}_{10} M_{*}/M_{\odot}\simeq10.5)$, this to guarantee that the sample was complete at stellar masses above those limits. The resulting sample contains 1,804 galaxies. 

\begin{figure}[h!tb]
\small
\centering
\begin{tabular}{@{}llll@{}}
\midrule
Telescope/Camera & Filter Name & $\underset{[\mu m]}{\lambda_{mean}} $ & $~~~~~~\underset{25\% / 50\% / 75\%}{\rm MAG_{AB}}$ \\ \midrule
GALEX            & FUV         & 0.154     &24.19 / 24.83 / 25.50
                                               \\
                 & NUV         & 0.231        &23.48 / 24.41 / 25.01
                                             \\
CFHT/Megacam     & U           & 0.382        &24.20 / 25.37 / 26.39
                                             \\
Subaru/HSC       & g           & 0.485        &23.46 / 24.69 / 25.80
                                             \\
                 & r           & 0.622        &22.57 / 23.78 / 24.93
                                             \\
                 & i           & 0.750        &21.74 / 22.91 / 24.18
                                             \\
                 & z           & 0.889        &21.29 / 22.30 / 23.51
                                             \\
                 & Y           & 0.976        &21.00 / 21.93 / 23.03
                                             \\
                 & IA427       & 0.427        &23.84 / 25.00 / 25.94
                                             \\
                 & IA464       & 0.464        &23.51 / 24.67 / 25.59
                                             \\
                 & IA484       & 0.484        &23.50 / 24.68 / 25.69
                                             \\
                 & IA505       & 0.505        &23.30 / 24.52 / 25.48
                                             \\
                 & IA527       & 0.527       &23.26 / 24.49 / 25.52
                                              \\
                 & IA574       & 0.574        &22.92 / 24.11 / 25.12
                                             \\
                 & IA624       & 0.624       &22.65 / 23.86 / 24.97
                                              \\
                 & IA679       & 0.679        &22.08 / 23.34 / 24.52
                                             \\
                 & IB709       & 0.709       &22.11 / 23.37 / 24.59
                                              \\
                 & IB738       & 0.738       &21.95 / 23.17 / 24.44
                                              \\
                 & IB767       & 0.767        &21.74 / 22.93 / 24.25
                                             \\
                 & IB827       & 0.827        &21.56 / 22.66 / 24.00
                                             \\
                 & NB711       & 0.711        &22.00 / 23.24 / 24.41
                                             \\
                 & NB816       & 0.816        &21.60 / 22.72 / 24.04
                                             \\
U-VISTA          & Y           & 1.021       &20.92 / 21.81 / 22.86
                                              \\
                 & J           & 1.253        &20.51 / 21.30 / 22.19
                                             \\
                 & H           & 1.645        &20.09 / 20.84 / 21.62
                                             \\
                 & Ks          & 2.153        &19.63 / 20.30 / 21.08
                                             \\
                 & NB118       & 1.191        &20.70 / 21.49 / 22.35
                                             \\
\textit{Spitzer}/IRAC             & CH1         & 0.744    &19.49 / 19.98 / 20.49
                                \\
                 & CH2         & 1.011             &19.59 / 20.02 / 20.43
                                        \\
                 & CH3         & 1.408               &19.63 / 20.04 / 20.41
                                      \\
                 & CH4         & 2.879               &19.75 / 20.24 / 20.68
                                      \\
\textit{Spitzer}/MIPS          &    --------      & 23.674      &17.43 / 17.93 / 18.50
                                               \\\bottomrule
\end{tabular}
\captionof{table}{Broad and narrow band set of filters employed as input data in the \cigale\ SED fitting, where $ \lambda_{mean}$ define the central wavelength of the filters and MAG$_{\rm AB}$ quantify the apparent magnitudes for each filter over the 0.25, 0.5, and 0.75 quantiles.}
\label{filters}
\end{figure}

\subsubsection{Radio based SFR computation} \label{radiosfr}

The 1.4\,GHz luminosity, $L_{1.4 \textrm{GHz}}$ of our galaxies was derived by \citet{Jimenez19} from the observed flux densities at 3\,GHz, $S_{3\textrm{GHz}}$, as :
\begin{equation}
    L_{1.4\textrm{GHz}}  = \frac{4\pi D_L^2(z)}{(1+z)^{1-\alpha}}\left(\frac{1.4}{3}\right)^{-\alpha} \frac{S_{3\textrm{GHz}},}{\textrm{W}~\textrm{Hz}^{-1}~\textrm{m}^{-2}} 
\end{equation}
\noindent with $D_L$ the luminosity distance in meters and $\alpha$ the spectral index of the synchrotron power law ($S_\nu \propto \nu^{-\alpha}$) \citep{Condon91} of 0.7, more adapted for the VLA COSMOS 3GHz sample \citep{smolcic17a}. Then \citet{Jimenez19} derived SFR estimates using the IR-radio calibrations as follow:
\begin{equation}
    \frac{\textrm{SFR}_{\textrm{radio}}}{M_\odot/yr}=f_{IMF}10^{-24}10^{q_{IR}}\frac{L_{1.4\textrm{Ghz}}}{\textrm{W}~\textrm{Hz}^{-1}} \label{sfr_rad}
\end{equation}
\noindent where $f_{IMF}$ depends on the initial mass function (IMF) (e.g. for a Salpeter IMF $f_{IMF}=1.72$), and $q_{IR}$ in the case of FIR detected star forming galaxies only as $q_{IR}=(2.83\pm 0.02)\times(1+z)^{-0.15\pm0.01}$ \citep{Delhaize17}.  As this calibration are based on the IR-SFR calibration, it relies on the hypothesis of a constant SFR in the last 100\,Myrs \citep{kennicutt12}.

\subsection{\textbf{Optical counterparts and photometric redshifts}}

We cross-matched the radio-selected sample from \citet{Jimenez19} with the latest photometric catalog released on the COSMOS field \citep{Weaver_2022}. We adopt \textsc{The farmer} version of the COSMOS2020 photometric catalog. The fluxes are estimated using a profile-fitting tool. Such a tool presents the advantage of being more robust against the blending of nearby sources \citep{Weaver_2022}. Moreover, the total flux is directly derived in all the bands, despite variations in the Point Spread Function (PSF). Such PSF variations are taken into account in the modeling of the light profile \citep{Lang16}, while uncertain corrections are required when using aperture fluxes \citep[][]{Laigle16}.

Sources were cross-matched in position using a maximum radius of 1.5$\arcsec$. We recovered 1584 sources, which is lower than the initial 1804 sources from the \citet[]{Jimenez19} sample. This difference is explained by the large masked areas around bright sources in COSMOS2020, and the difficulty for \textsc{The farmer} source extraction method to perform a profile-fitting around the brightest sources \citep{Weaver_2022}. The photometry is extracted in the 32 bands listed in Table~\ref{filters}. A signal-to-noise (SNR) threshold larger than 3 has been applied in the \textit{Spitzer}/MIPS 24\,$\mu$m band. We also provide the quantiles of the magnitude distribution in each band to characterize the brightness of our sample.

\begin{figure}[h!]
    \includegraphics[width=\columnwidth]{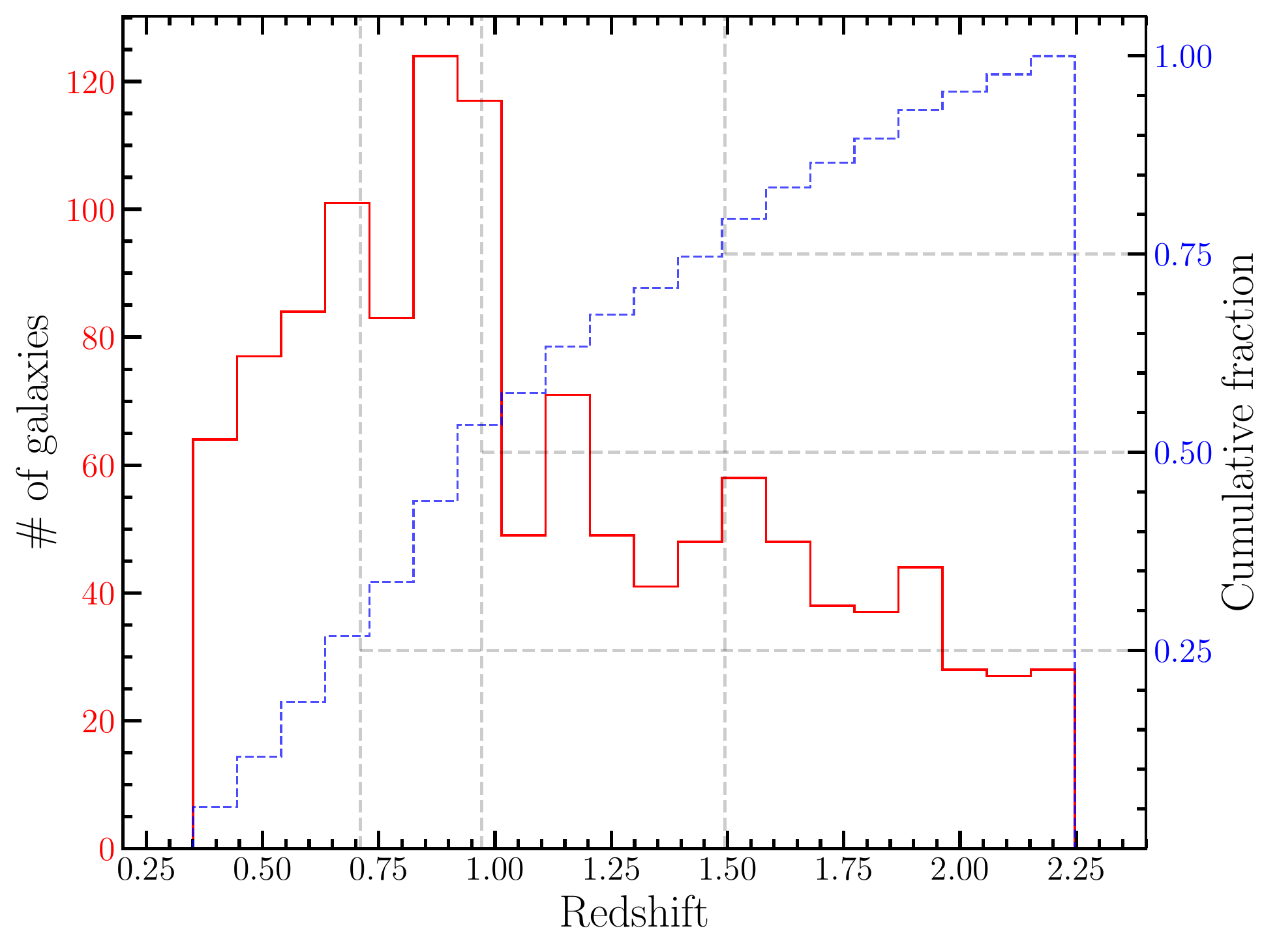}
 \caption{ Redshift distribution of the final sample, extracted from the COSMOS VLA 3 GHz survey
project which results from the different criteria selection applied in \citet{Jimenez19}. Cross-match with \textsc{The farmer} version of the COSMOS2020 \citep{Weaver_2022} for photometric measures leads to a mass complete sample of 1584 star-forming galaxies over the redshift range $0.35<z<2.25$. Dashed gray lines indicate the 0.25, 0.5 and the 0.75 quantiles.}
 \label{z-dist}
 \vspace{10pt}
\end{figure}

Finally, we keep the same photometric redshifts as \citet{Jimenez19}, in order to be consistent with their radio SFRs derivation and AGN removal. These photometric redshifts have been derived with the template-fitting code \texttt{Le Phare} \citep{Arnouts02, Ilbert06} on the COSMOS2015 catalog \citep{Laigle16}. We compare the photometric redshifts with the spectroscopic sample available in COSMOS and described in Sect. 2.9 of \citet{Weaver_2022}. We find a precision of $\sigma_{\Delta z / (1+z_s)}$= 0.024 and an outlier fraction of $\eta$ = 1\%. Figure~\ref{z-dist} shows the final redshift distribution of our 1584 selected sources, with a mean photometric redshift of $z\simeq 1$.

\begin{table*}
    \begin{center}
    \begin{tabular}{lll}
    Parameter     & Value           &                        \\ 
    \hline
    \hline
    \multicolumn{3}{c}{Star formation history}                   \\
    \hline
    \multicolumn{3}{c}{Non-parametric: \texttt{sfhNlevels}}                        \\ 
    \hline
    $age$ {[}Gyr{]} & {[}2.8;9.8{]} & $age$ of Universe at galaxies' redshift ; 10 values linearly sampled\\
    N$_{SFH}$& 1000      & $\#$ of SFHs for each $age$ value                                  \\
    \hline
    \multicolumn{3}{c}{$\tau$-delayed SFH + flexibility}\\ 
    \hline
    $age$ [Gyr] &    [2.8;9.9] & Age of the galaxy ; 10 values linearly sampled                               \\
    $\tau_{main}$ [Gyr]&   [0.1;20] & 10 values linearly sampled                                 \\
    $age_{trunc}$ [Myr]& 30, 100, 300, 500 &  Age of the star-formation burst or quenching event \\
    $r_{\rm SFR}$& $[10^{-3};10^{3}]$ & Strength of the burst or quenching. 10 values log-sampled \\ 
    \hline
    \hline
    \multicolumn{3}{c}{Dust emission: \citep{Dale14}}\\ 
    \hline
    $\alpha$ & 1.5, 2.0, 2.5  &  far-IR slope                                     \\ 
    \hline
    \hline
    \multicolumn{3}{c}{Dust attenuation: \citep{Calzetti00} }                        \\ 
    \hline
    $E(B-V)s$ & [0,0.8] & 10 values linearly sampled  \\ 
    \hline
    \hline
    \multicolumn{3}{c}{AGN activity: \texttt{SKIRTOR} \citep{Stalevski16} }                        \\ 
    \hline 
    $\textrm{frac}_{\textrm{AGN}}$ & [0,0.5] & Contribution of the AGN to the total L$_{IR}$ ; 10 values linearly sampled \\ \hline
    \end{tabular}
    \end{center}
    \caption{ Parameters of the models used to perform the SED modeling with \cigale. We refer the reader to \cite{Boquien19}, \cite{Ciesla17}, and \citet{Ciesla22} for more details on the \cigale\ input parameters. }
\label{cig_tab}
\end{table*}

\section{SED modelling with \cigale\ } \label{s-m}

To model the spectral energy distribution of our galaxy sample, we use the SED modeling code \cigale\footnote{\url{https://cigale.lam.fr/}}  \citep{Boquien19}. \cigale\ can builds and fits physical models from X-ray to radio, taking into account the energy budget between the light absorbed in the UV-optical and re-emitted in IR by dust. 
It uses a Bayesian-like analysis to derive the physical properties of galaxies.
Its versatility is characterized by the multiple modules modeling the SFH of galaxies, the stellar, dust , and nebular emission, the AGN contribution, as well as the radio emission of galaxies.
In particular, the SFH can be handled through analytic, non-parametric, and simulated SFHs \citep{Boquien14, Ciesla15, Ciesla17}.
In this work, we use the recently added non-parametric SFH module \texttt{sfhNlevels} \citep{Ciesla22}, \citet{BruzualCharlot03} stellar population models, a \citet{Calzetti00} attenuation law, and \citet{Dale14} dust emission library.
For the purpose of testing the robustness of our results in Sect.~\ref{SFH_model_check}, we will also consider a classical parametric SFH using a $\tau$-delayed model plus flexibility \citep{Ciesla17}. For the standard parametrisation of the code, we do not include the contribution of an AGN, except for a test presented in Sect.~\ref{AGN_model_ceck}.
Table~\ref{cig_tab} presents the parameters used to set up the \cigale\ fitting procedure. 
 The accuracy of the physical parameters extracted from SED-fitting benefits from the depth of the imaging data and the multi-wavelength coverage from FUV (GALEX at 0.15\,$\mu$m) to mid-IR (\textit{Spitzer}/MIPS at 24\,$\mu$m) presented in Table~\ref{cig_tab}. 
 
A probability distribution function is associated with each parameter estimated with CIGALE. The flux uncertainties are handled within the $\chi^2$ estimate. An additional of 10\% is included by \cigale\ \citep[see][]{Boquien19}. The final flux uncertainty considered in the $\chi^2$ estimate is given by err$_{tot}$=$\sqrt{\textrm{err}_{\textrm{flux}}^2 + (\textrm{flux}\times0.1)^2}$.

\begin{figure*}
     \includegraphics[width=1\textwidth]{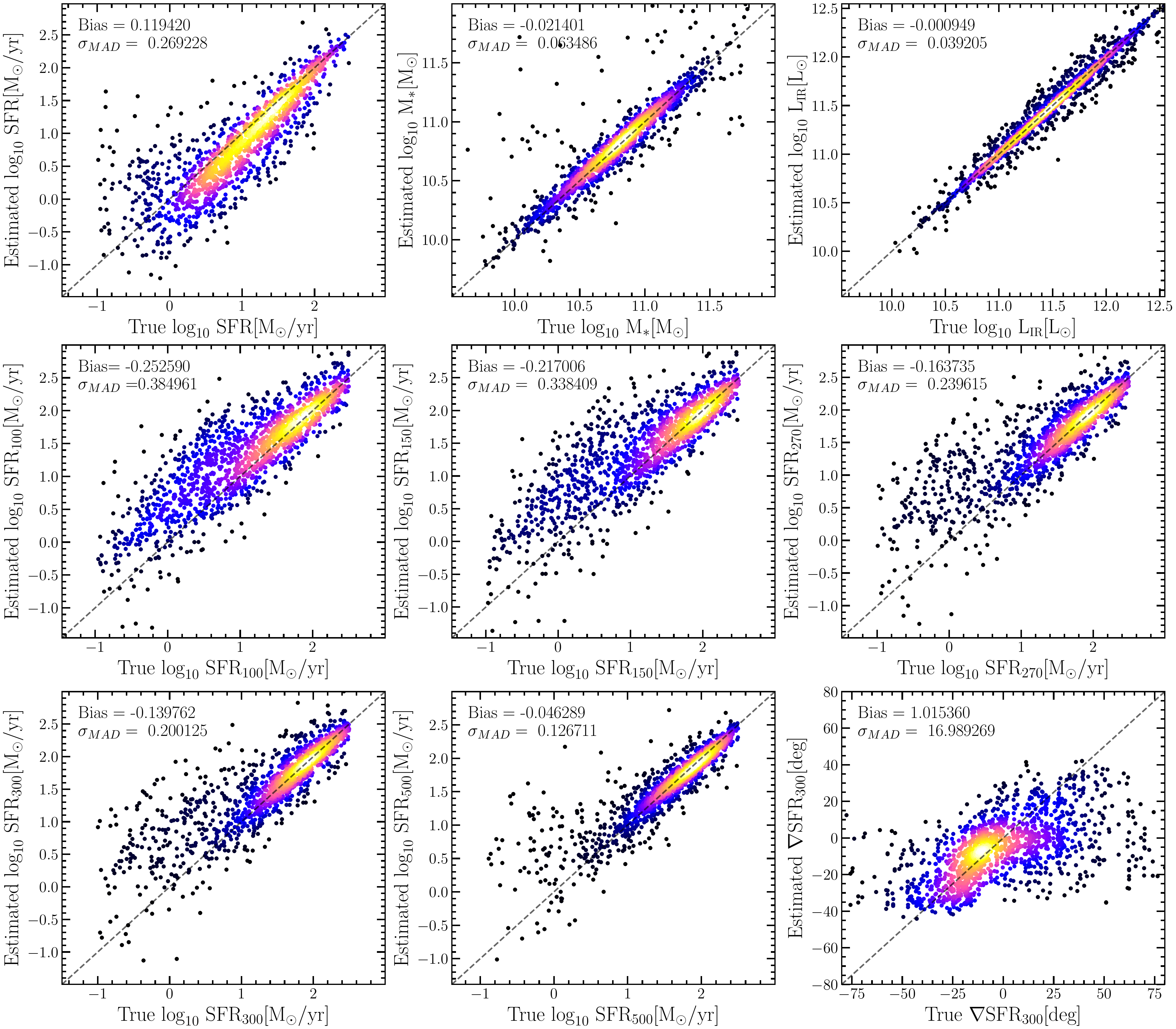}
        \caption{ Results of the mock analysis for the SED modeling using the non-parametric SFH model. The input parameters used to build the mock catalog are shown on the $x$-axis while the results of the fitting of the mock catalogs are shown on the y-axis. From left to right: upper present the instantaneous SFR, the stellar mass M$_*$ and the infrared luminosity L$_{\textrm{IR}}$; mid panels present the time averaged SFRs for the last 100, 150, and 270\,Myr over the SFH, and bottom panels present the time averaged SFRs for the last 270 and 500\,Myr over the SFH as well as the SFR gradient computed over the last 300\,Myr. The one-to-one relationship is indicated by the black solid lines. The bias and precision estimated for the parameter are indicated in each panel.}
        \label{mock_N-levels}

\end{figure*}

\subsection{\label{non-param} Non-parametric SFH: the \texttt{sfhNlevels} module}

In this \cigale\ module, the SFH is modeled using a given number of bins with constant SFR.
The SFRs of two consecutive bins is compelled by a given prior.
In this work, we use a \textit{continuity} prior following a \textit{Student-t} distribution \citep{Leja19}.
A given number of SFH (N$_{\rm SFH}$) is computed by randomly selecting the SFR in each bin from the distribution imposed by the prior.

This module includes the computation of a parameter called the SFR gradient which characterizes the evolution of a galaxy over the SFR-M$_*$ plane. This parameter is defined as an angle (in degrees) and describes the direction of evolution of a galaxy in a given time interval $\Delta t= t_2-t_1$ where the SFR changes by $\Delta \textrm{log}_{10}$SFR and the mass by $\Delta \textrm{log}_{10}$M$_*$ :
\begin{equation}
    \nabla \textrm{SFR}_{\Delta t} = \arctan \left(\frac{\Delta \textrm{log}_{10}\textrm{SFR}}{\Delta \textrm{log}_{10}\textrm{M}_*}\right).
\end{equation}

With this definition, galaxies with $\nabla \textrm{SFR}_{\Delta t}<0$ present a declining star-formation activity while galaxies with $\nabla \textrm{SFR}_{\Delta t}>0$ an enhanced one, over the last time interval $\Delta t$. 
This new module of \cigale\ and the SFR gradient parameter is presented and tested in \citet{Ciesla22}.

\subsection{Mock analysis}\label{sec:mock}

To assess the reliability of the parameters derived with \cigale, we build a mock catalog mimicking the data sample. For each galaxy of the sample, we simulate mock fluxes from the best fit of their observed fluxes: we start with the flux densities obtained by integrating the best-fit template over the same set of filters as the original one. These mock flux densities are then perturbed by adding a noise randomly selected in a Gaussian distribution with a standard deviation $\sigma$ corresponding to the error of the original flux density. By running the code on this synthetic sample, for which all the physical parameters are known, we can compare the exact values of the physical parameters and the retrieved ones.  This test provides an evaluation of the ability of the code to provide constrained values of the physical parameters we want to study \citep[for instance,][]{Noll09,Buat14,Ciesla15,Boquien19}.
Fig.~\ref{mock_N-levels} shows the results of the mock analysis for our SFG sample. 
The stellar mass, SFR, and IR luminosity parameters are well recovered, which is expected given that the wavelength coverage of the sample allows to probe the UV rest-frame, NIR rest-frame, as well as a data point to constrain the mid-IR \citep[see for instance,][]{Buat14,Ciesla15,Malek18}.
For the purpose of this study, we analyse the results of the mock analysis of a few SFH parameters as well.
For the SFR gradient mentioned is Sect.~\ref{non-param}, computed over 300\,Myr, there is a overall good agreement between the true values of the SFR gradient and the ones derived from \cigale\ (Fig.~\ref{mock_N-levels}, bottom right panel).
This choice of time interval provides a fair trade between a good indicator of the recent SFH and a well-constrained parameter \citep[see the tests performed in][]{Ciesla22}. 

Finally, we also check the reliability of the SFR estimates averaged over a number of different time intervals that we will use in the analysis in the following sections.
These estimates are well constrained. Indeed, the bias and the precision are generally in the range 0.2-0.3 dex, as indicated in Fig.~\ref{mock_N-levels}. 

\section{Results} \label{res}
\begin{figure}
  \begin{center}
    \includegraphics[width=\columnwidth]{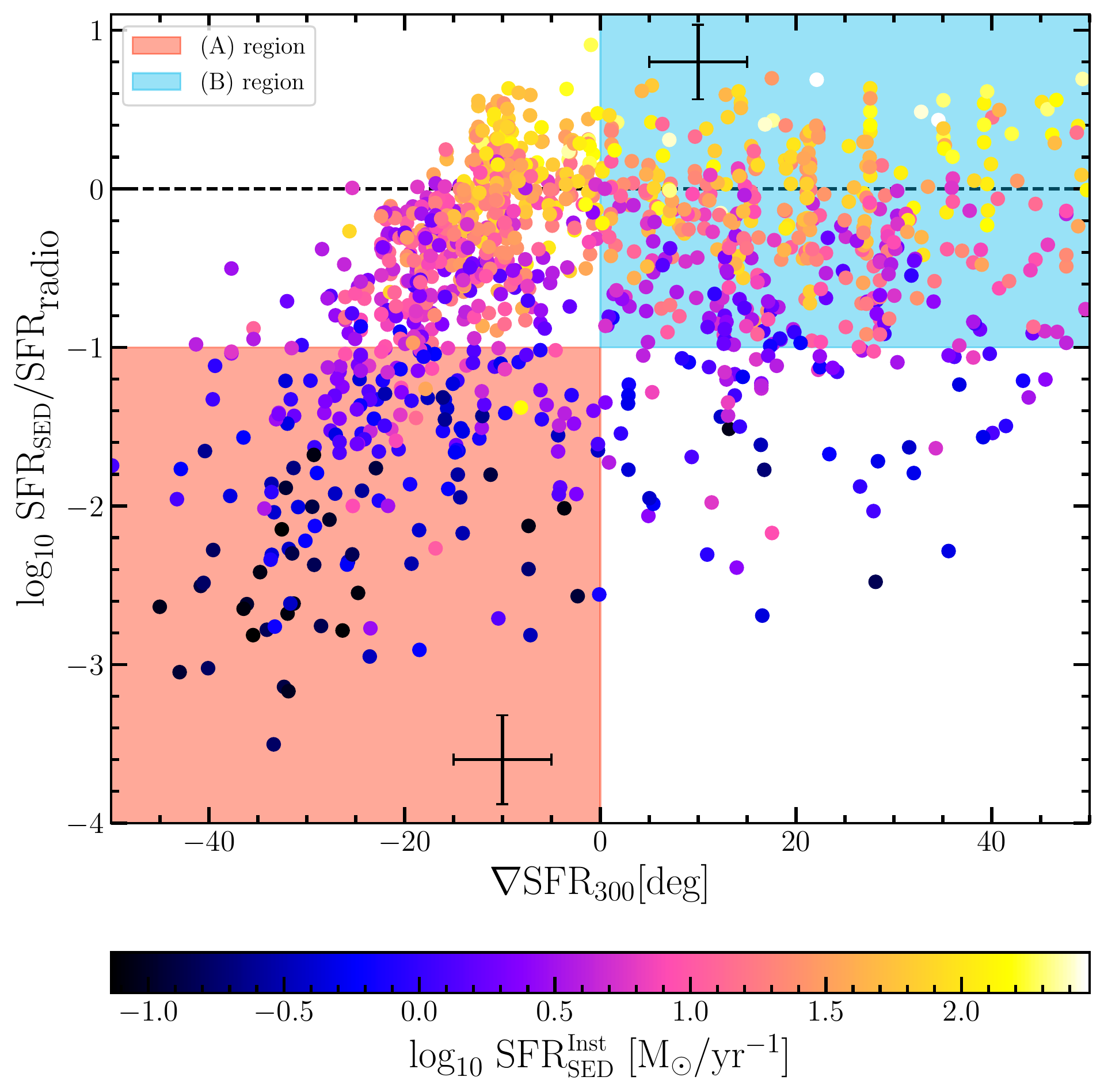}%
  \end{center}
        \caption{ Ratio between  $\textrm{SFR}_{\textrm{SED}}^{\textrm{Inst}}$ and SFR$_{\textrm{radio}}$ as a function of the SFR gradient over the last 300\,Myr ($\nabla \textrm{SFR}_{300}$). In a perfect scenarios where the two indicators give the same value, one expects a uniform distribution centered at 1 (but actually 0 in log) over the SFRs ratios along the $\nabla \textrm{SFR}_{300}$. The red shaded region count for galaxies with negative values on $\nabla \textrm{SFR}_{300}$ (i.e., being quenched) and a SFRs ratio smaller than 0.1, while the blue shaded region count for galaxies with positive values on $\nabla \textrm{SFR}_{300}$ (i.e., being starbursting) and a SFR ratio above 0.1. Dots are color coded according to the instantaneous SFR derived from the SED fitting procedure. Mean error bars are shown in black crosses for galaxies inside the (A) and (B) regions.  }
        \label{scatter_chi}
\end{figure}

\begin{figure}[h!]
  \begin{center}
    \includegraphics[width=\columnwidth]{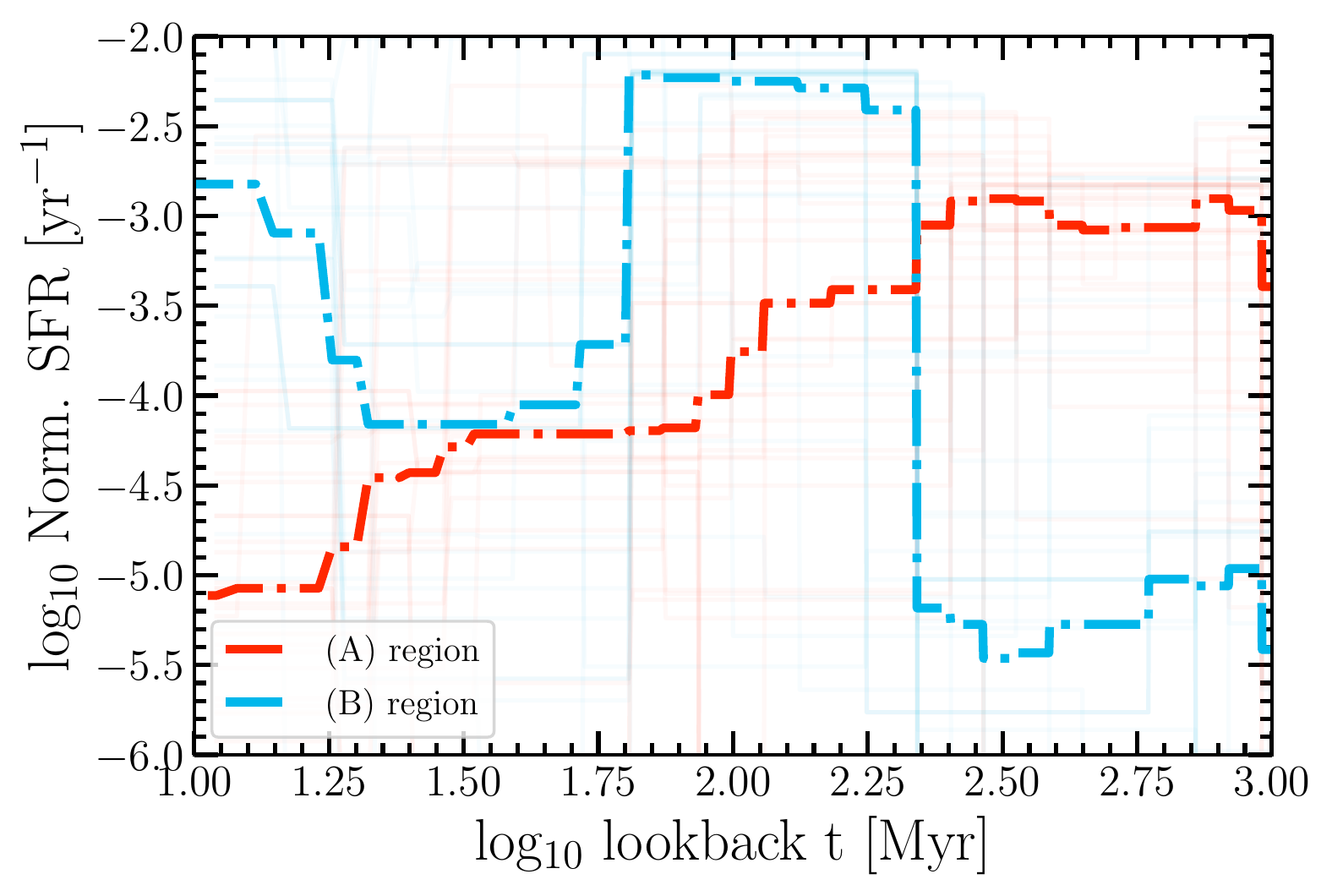}%
  \end{center}
        \caption{Mean normalized SFHs $\left(SFR(t)/\int SFR(t)\,dt\right)$ over 100 randomly selected galaxies for each A and B regions on Fig. \ref{scatter_chi}. In this representation we can clearly see the behavior on the recent SFH of these two populations defined by the $\nabla \textrm{SFR}_{300}$ parameters where galaxies on region (A) and (B) will present in average, an increase and decrease respectively in the recent SHF. This trend is present in the last 300\,Myrs but is more pronounced on the last 20\,Myrs on both galaxy present on the (A) and (B) regions.}
        \label{sfh_populations}
\end{figure} 

In this section, we compare the SFR obtained from radio observations by \citet{Jimenez19} ($\textrm{SFR}_{\textrm{radio}}$) with the instantaneous SFR obtained from SED fitting ($\textrm{SFR}_{\textrm{SED}}^{\rm inst}$) with \cigale{}.

Fig.~\ref{scatter_chi} shows the ratio $\textrm{SFR}_{\textrm{SED}}^{\rm inst}/\textrm{SFR}_{\textrm{radio}}$ as a function of the gradient $\nabla \textrm{SFR}_{300}$ computed over the last 300\,Myr.  A positive value of $\nabla \textrm{SFR}_{300}$ indicates a sustained star-formation activity while a large negative gradient indicates a quenching phase. If both SFR$_\textrm{SED}^{\rm inst}$ and
SFR$_\textrm{radio}$ estimators were unbiased and tracing the instantaneous SFR,  we would expect a distribution centered on 1 (0 in log) in Fig.~\ref{scatter_chi}, independently of the $\nabla \textrm{SFR}_{300}$. However, galaxies with values of $\nabla \textrm{SFR}_{300}\leq0$ are spread over almost four magnitudes, while galaxies with $\nabla \textrm{SFR}_{300}\gtrsim0$ show a better agreement between both SFR indicators. 

To quantify this trend, we define two regions in Fig.~\ref{scatter_chi}, with the red shaded region (A) having $\nabla \textrm{SFR}_{300}\leq0$ and $\textrm{log}_{10}~\textrm{SFR}_{\textrm{SED}}^{\rm inst}/\textrm{SFR}_{\textrm{radio}}<-1$, and the blue shaded region (B) having $\nabla \textrm{SFR}_{300}>0$ and $\textrm{log}_{10}~\textrm{SFR}_{\textrm{SED}}^{\rm inst}/\textrm{SFR}_{\textrm{radio}}>-1$. 
 The regions (A) and in (B) represents 
$\sim$19$\%$ and $\sim$31$\%$ of our radio galaxy sample, respectively.
We observe in Fig.~\ref{scatter_chi} that the ratio $\textrm{SFR}_{\textrm{SED}}^{\rm inst}/\textrm{SFR}_{\textrm{radio}}$ drops continuously as the gradient decreases below zero. By selecting galaxies at $\nabla \textrm{SFR}_{300}\leq -20$, we find that $\sim 75\%$ of them have a SFR ratio which differs by more than a factor 10 (i.e. $\textrm{log}_{10}~\textrm{SFR}_{\textrm{SED}}^{\rm inst}/\textrm{SFR}_{\textrm{radio}}<-1$). By contrast, the SFR differs by $>10$ for only $\sim 25\%$ when we consider galaxies at $\nabla \textrm{SFR}_{300}>0$.

\begin{figure*}[h!]
\centering
\subfloat[]{ \includegraphics[width=0.33\textwidth]{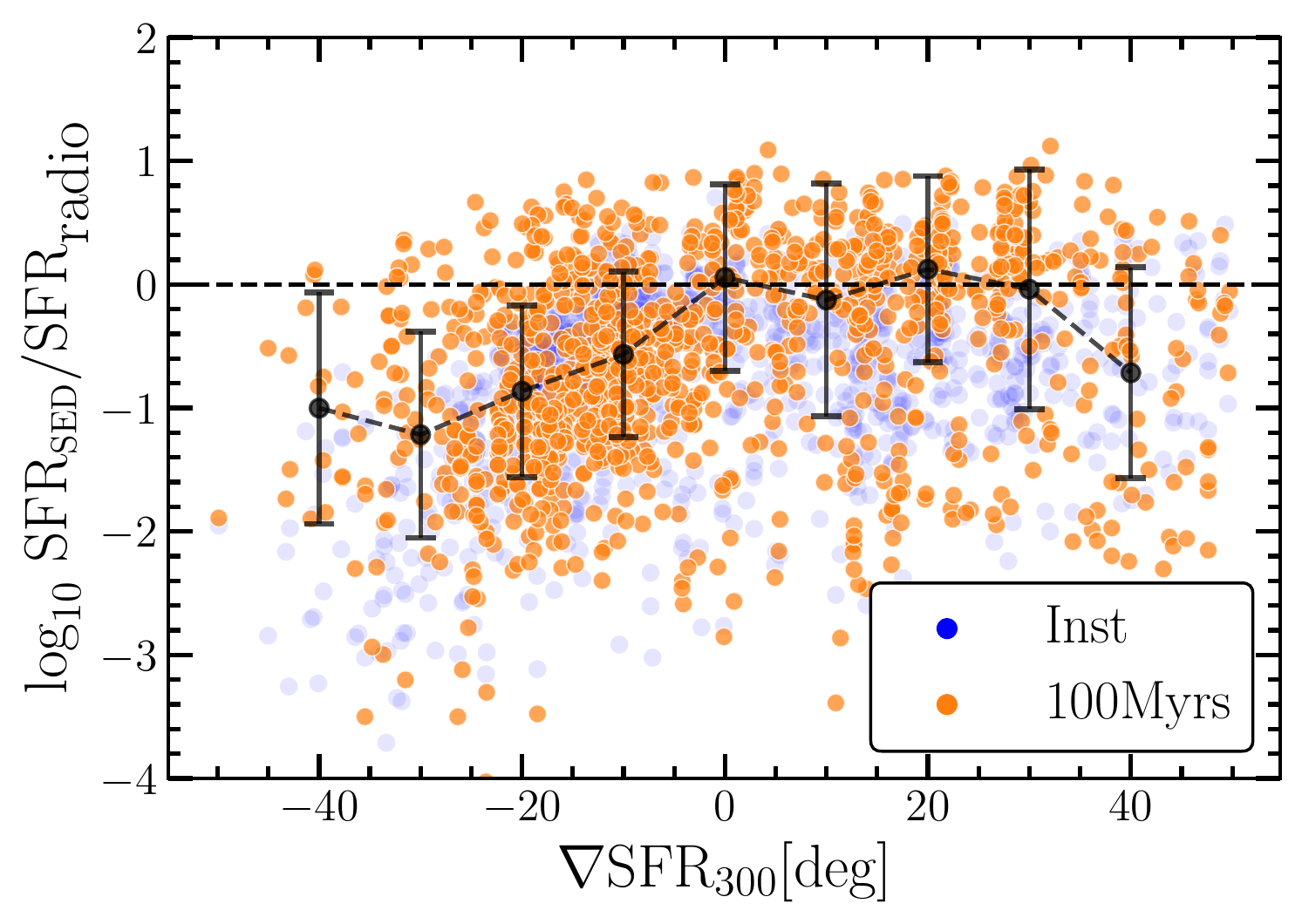}}
\subfloat[]{ \includegraphics[width=0.33\textwidth]{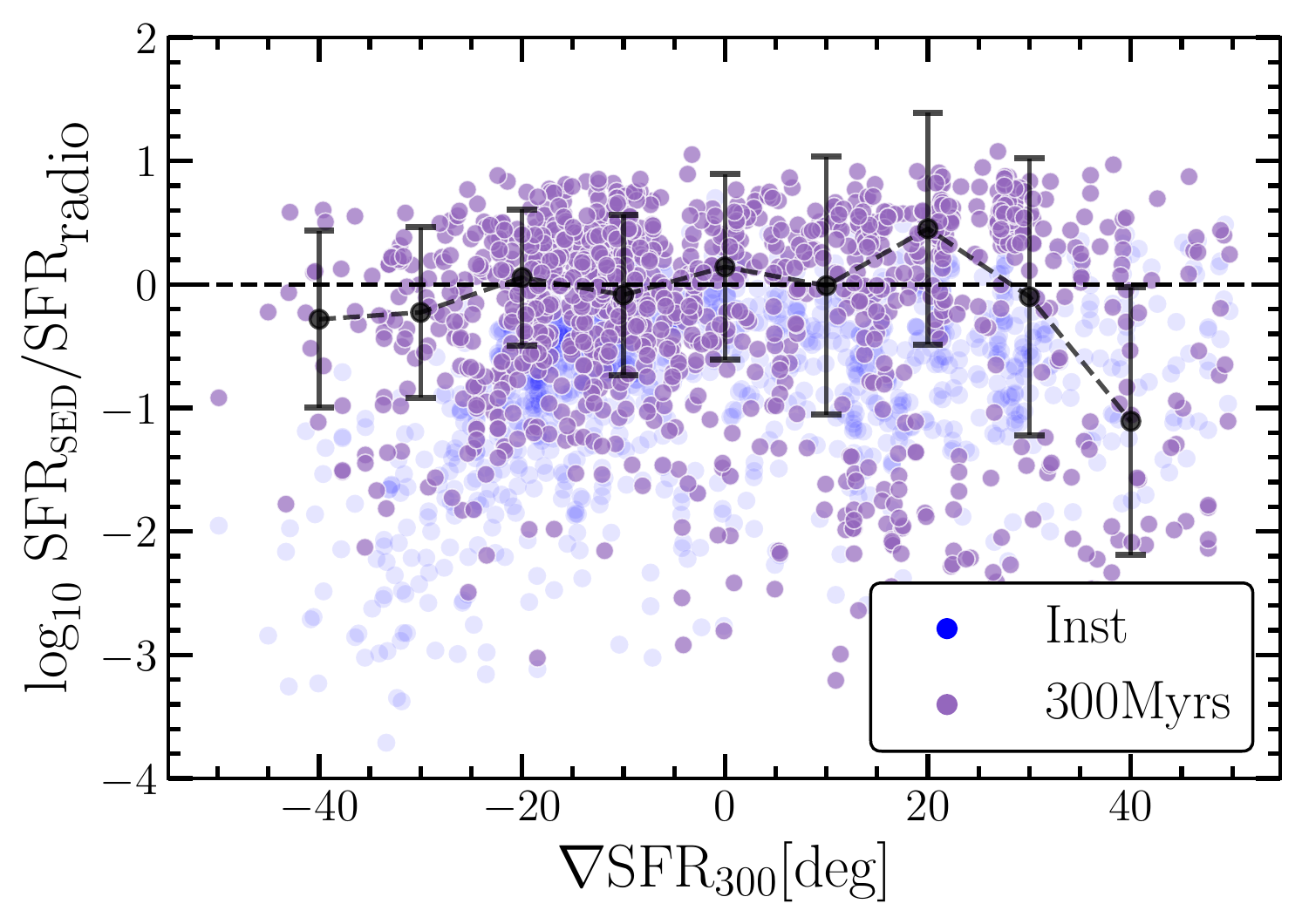}}
\subfloat[]{ \includegraphics[width=0.33\textwidth]{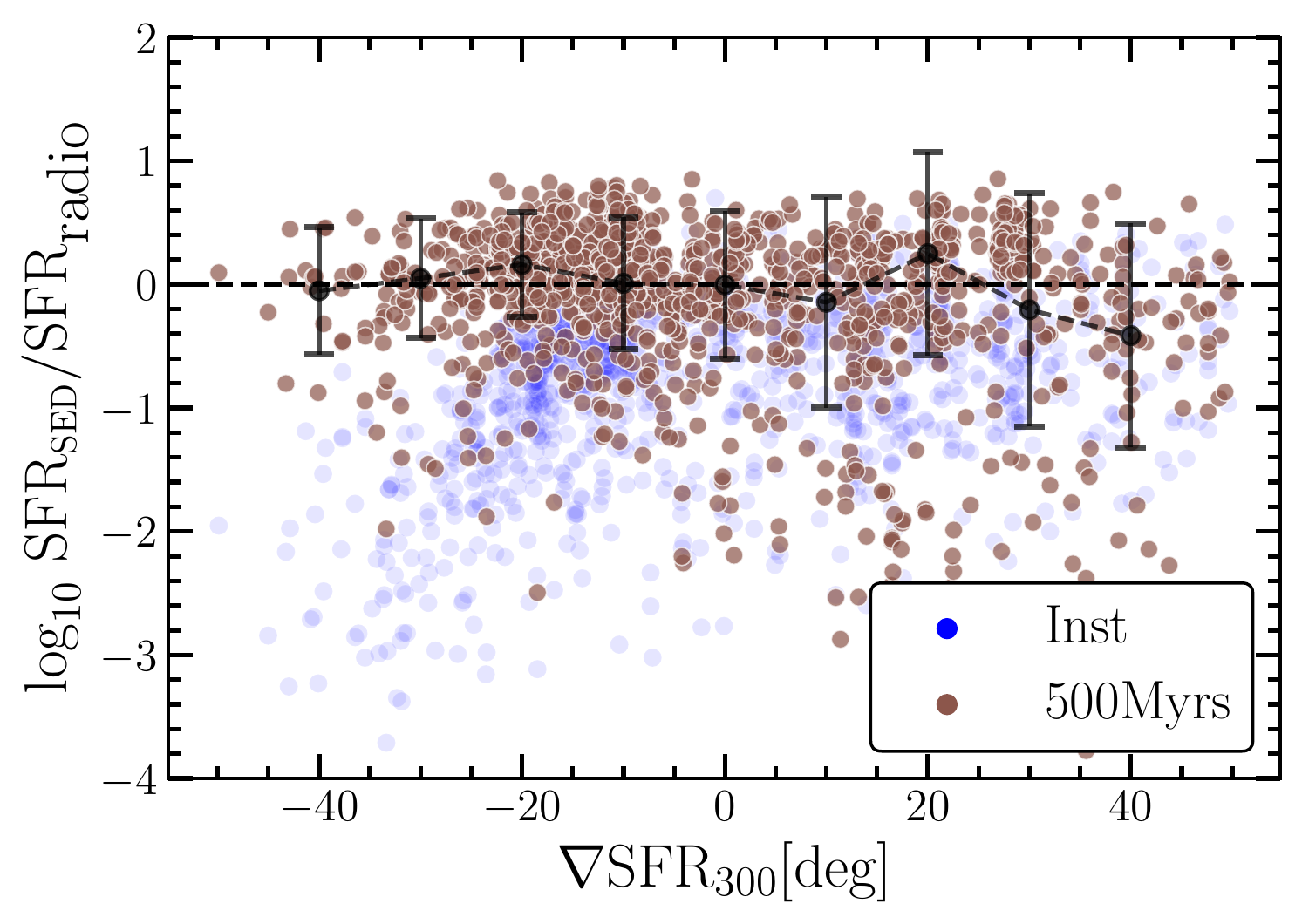}}
\caption{    Ratios between the $\textrm{SFR}_{\textrm{ SED}}$ and the $\textrm{SFR}_{\textrm{radio}}$ as a function of the SFR gradient over the last 300\,Myr ($\nabla \textrm{SFR}_{300}$). Galaxies with $\nabla \textrm{SFR}_{300}<0$ present a recent quenching while galaxies with $\nabla \textrm{SFR}_{300}>0$ present a recent starburst. Blue dots for instantaneous values of SFR$_{\rm SED}^{\textrm{Inst}}$ while in the case of the orange, purple, and brown circles the SFR$_{\textrm{SED}}$ was averaged for last 100, 300, and 500\,Myr, respectively. Black dots shows the mean trend and the dispersion of SFR ratio along the $\nabla \textrm{SFR}_{300}$ inside the fixed bins size of 10 degrees. Color codes are shared with Fig.~\ref{ratios_distributions}.}
\label{3figs}
\end{figure*}

\begin{figure}[h!]
        \centering
 \includegraphics[width=\columnwidth]{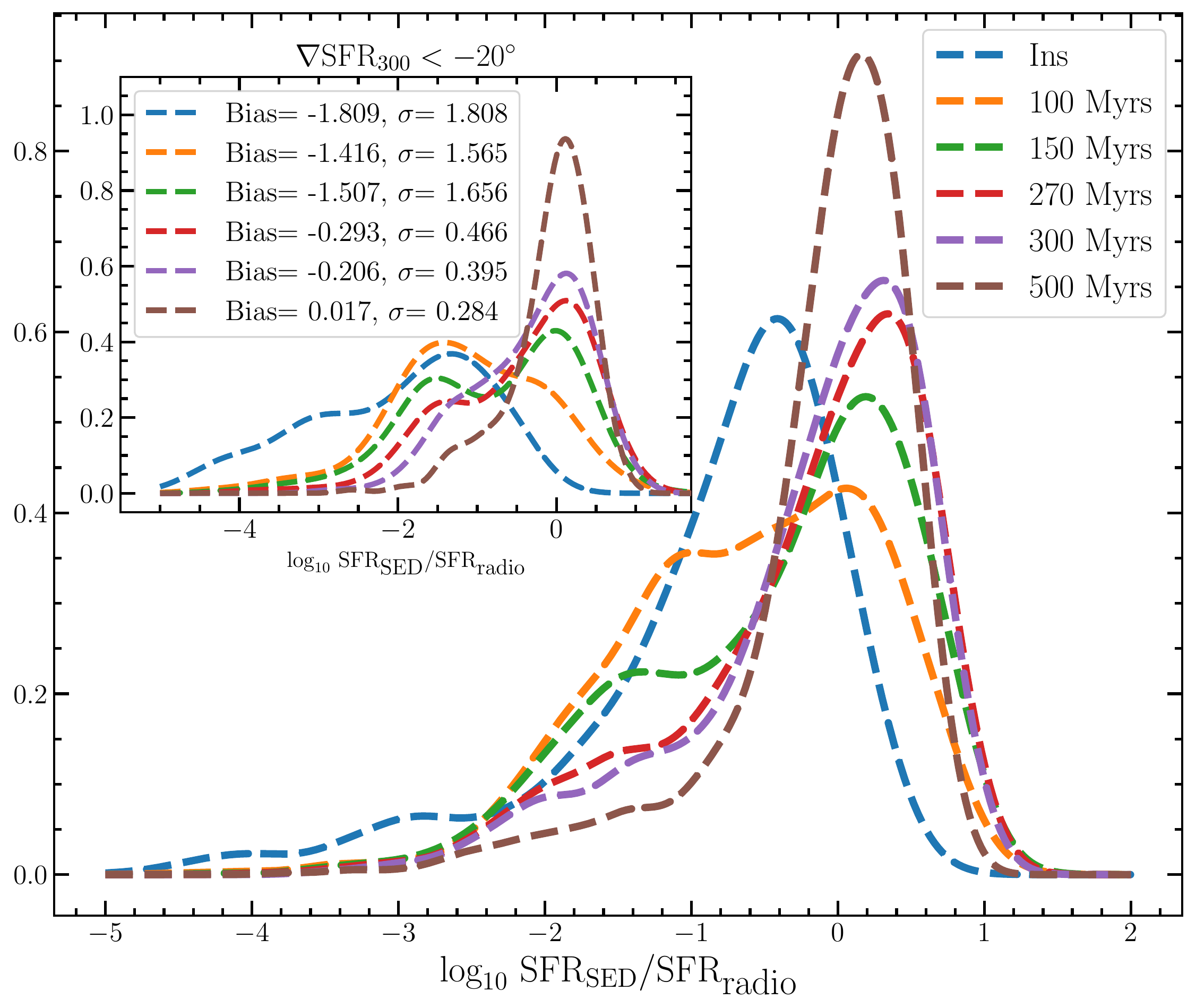}
        \caption{Distribution (in logarithmic scale) of the ratio between the radio and SED SFRs. Each color corresponds to $\textrm{SFR}_{\textrm{SED}}$ obtained by averaging the SFH over different timescales, going from 100\,Myr up to 500\,Myr. The inset presents the same statistic, but selecting only galaxies with a declining SFH ($\nabla \textrm{SFR}_{300}\leq -20$). The bias and the dispersion are indicated for each timescale.   }
        \label{ratios_distributions}
\end{figure}

In Fig.~\ref{sfh_populations}, we show the mean SFH of 100 randomly selected galaxies in these (A) and (B) shaded regions. In this representation, we clearly see that the recent SFH of these two populations evolve differently at recent time. The galaxies of the region A (B) present an increase with time (decrease respectively) of the recent star-formation, consistently with the definition of the $\nabla \textrm{SFR}_{300}$ parameter. This trend is present in the last 300\,Myr but is more pronounced on the last 20\,Myr. Additionally the stellar masses are quite different between the two regions, with massive and low specific SFR (sSFR) galaxies in region (A) (with a median of ${\rm log}_{10} {\rm sSFR} = -1.72$ and ${\rm log}_{10} {\rm M}_* = 11.03$, having sSFR in Gyr$^{-1}$ and stellar mass in $\Msol$), and more active galaxies in region (B) (with a median of ${\rm log}_{10} {\rm sSFR}=0.2$ and  ${\rm log}_{10} {\rm M}_* = 10.86$).

We checked that the $\chi^2_{red}$ distribution peaks at $\simeq$ 1, with no particular trend over the SFR ratios nor $\nabla \textrm{SFR}_{300}$. We conclude that poor quality fits do not explain the low SFR ratio seen for declining SFH. 

A possible interpretation is that both SFR estimators do not trace the SFR over the same timescale \citep[e.g.,][]{Schleicher13}. In this study, we consider the SFR derived from SED modelling as an instantaneous indicator, SFR$_{\textrm{SED}}^{\textrm{Inst}}$. Using the individual SFH produced with \cigale\, we can artificially construct SFR$_{\textrm{SED}}$ indicators sensitive over longer timescales. To do so, we average the galaxy's SFH over different timescales $\Delta t$ (100, 150, 270, 300, and 500\,Myr) and compare them with the SFR$_{\textrm{radio}}$. Hereafter we note $\textrm{SFR}_{\rm \Delta t}$ the SFH averaged over a time interval $\Delta t$. Fig.~\ref{3figs} presents the ratio between SFR$_{\textrm{radio}}$ and SFR$_{\Delta t}$ as a function of the SFR gradient $\nabla \textrm{SFR}_{300}$ for three different timescales 100, 300 and 500\,Myr (in addition to the instantaneous SFR$_{\textrm{SED}}^{\textrm{Inst}}$ shown in blue). 
In Fig.~\ref{ratios_distributions}, we show the distribution of the SFRs ratios for all the different timescales mentioned above. We find that the agreement between both SFR indicators improves continuously as we consider a longer timescale to integrate the SFH. The bias decreases drastically when considering a timescale $>150$\,Myr, with a bias falling below 0.3 dex for the population with $\nabla \textrm{SFR}_{300}\leq -20$ at a timescale of 270\,Myr. If we consider a timescale of 270 Myr to estimate SFR$_{\textrm{SED}}$, we find that $\sim4\%$ of the SFRs differ by a factor $>$10 when $\nabla \textrm{SFR}_{300}\leq -20$ (to be compared with the 75\% considering an instantaneous SFR$_{\textrm{SED}}^{\textrm{Inst}}$). We conclude that the radio and SED-fitting SFR tracers converge toward a consistent value if SFR$_{\textrm{SED}}$ is averaged over timescales longer than 150\,Myr. The SFR$_{\textrm{radio}}$ and SFR$_{\textrm{SED}}$ of sources with a SFH with a flat or positive gradients are already in excellent agreement for a timescale of 100\,Myr. 

To summarize, there is an effect of the SFH on the discrepancy between the two SFR indicators. Galaxies that are the most affected by the discrepancy are those with $\nabla \textrm{SFR}_{300}\leq -20$, i.e. galaxies where star formation stops rapidly, quenching the star formation \citep{Ciesla22}. 

\section{Additional considerations}

In this section, we present additional analysis to test the robustness of the results presented in Sect.~\ref{res}, exploring several factors which could explain the discrepancy between the two SFR indicators in galaxies with decreasing star formation activity.

\subsection{Malmquist bias due to the radio flux selection}

Because of the flux selection, we can only select galaxies above a given radio SFR, approximately 10 $\Msolperyr{}$. The SFR derived from SED-fitting does not suffer from the same limitation. As a consequence, the ratio distribution is skewed toward lower values by construction. Indeed, SFR$_{\textrm{SED}}^{\textrm{Inst}}$ can take low values while $\SFRradio{}$ can not because of radio flux selection. Moreover, low values of the gradient $\gradient{}$ correspond to low values of SFR$_{\textrm{SED}}^{\textrm{Inst}}$, as shown in Fig.~\ref{scatter_chi}. Therefore, above a given level of uncertainties in SFR$_{\textrm{SED}}^{\textrm{Inst}}$ (especially at low gradient), a Malmquist bias could explained the observed trend.

We use the mock catalog presented in Sect.~\ref{sec:mock} to test such assumption. We reproduce a figure similar to Fig.~\ref{scatter_chi}, by studying the ratio between the estimated and true  SFR$_{\textrm{SED}}^{\textrm{Inst}}$ as a function of the gradient in the simulation. We selected simulated galaxies with their true SFR$_{\textrm{SED}}^{\textrm{Inst}}$ above 10 $\Msolperyr{}$, to mimic the flux selection of the radio sample. We find that the noise expected in the true SFR$_{\textrm{SED}}^{\textrm{Inst}}$ is not sufficient to create a trend such as the one seen in real data. We do not detect the trend in the simulation for any considered timescale. 
Moreover, the trend disappears when sufficiently long timescales are considered to estimate SFR$_{\Delta t}$, while the expected precision on SFR$_{\textrm{SED}}^{\textrm{Inst}}$ remains similar according to the simulation presented in Sect.~\ref{sec:mock}.
Only a bias of SFR$_{\textrm{SED}}^{\textrm{Inst}}$ toward lower values and a low gradient could explain the trend observed in the data. If any bias, we expect it in the opposite direction according to Fig.~\ref{mock_N-levels}.

\subsection{Impact of the assumed SFH} \label{SFH_model_check}

To test if the results presented in Sect.~\ref{res} are sensitive to the assumed SFH model, we fit again the whole sample with \cigale\ but using a flexible $\tau$-delayed SFH, as presented in \citet{Ciesla17}. This SFH has been proposed to disconnect the SFR and stellar mass estimates thanks to a flexibility in the recent SFH. Indeed, it allows for an instantaneous and recent starburst or rapid quenching. The time when this happens is a free parameter, as well as the intensity of the burst/quenching. This parametric SFH provides good estimates of the physical properties of galaxies, especially in terms of SFR \citep{Ciesla17,Schreiber18, Ciesla18}. The input parameters used to generate these SFH are provided in Table~\ref{cig_tab}. 

\begin{figure}[ht]
    \centering
    \includegraphics[width=\columnwidth]{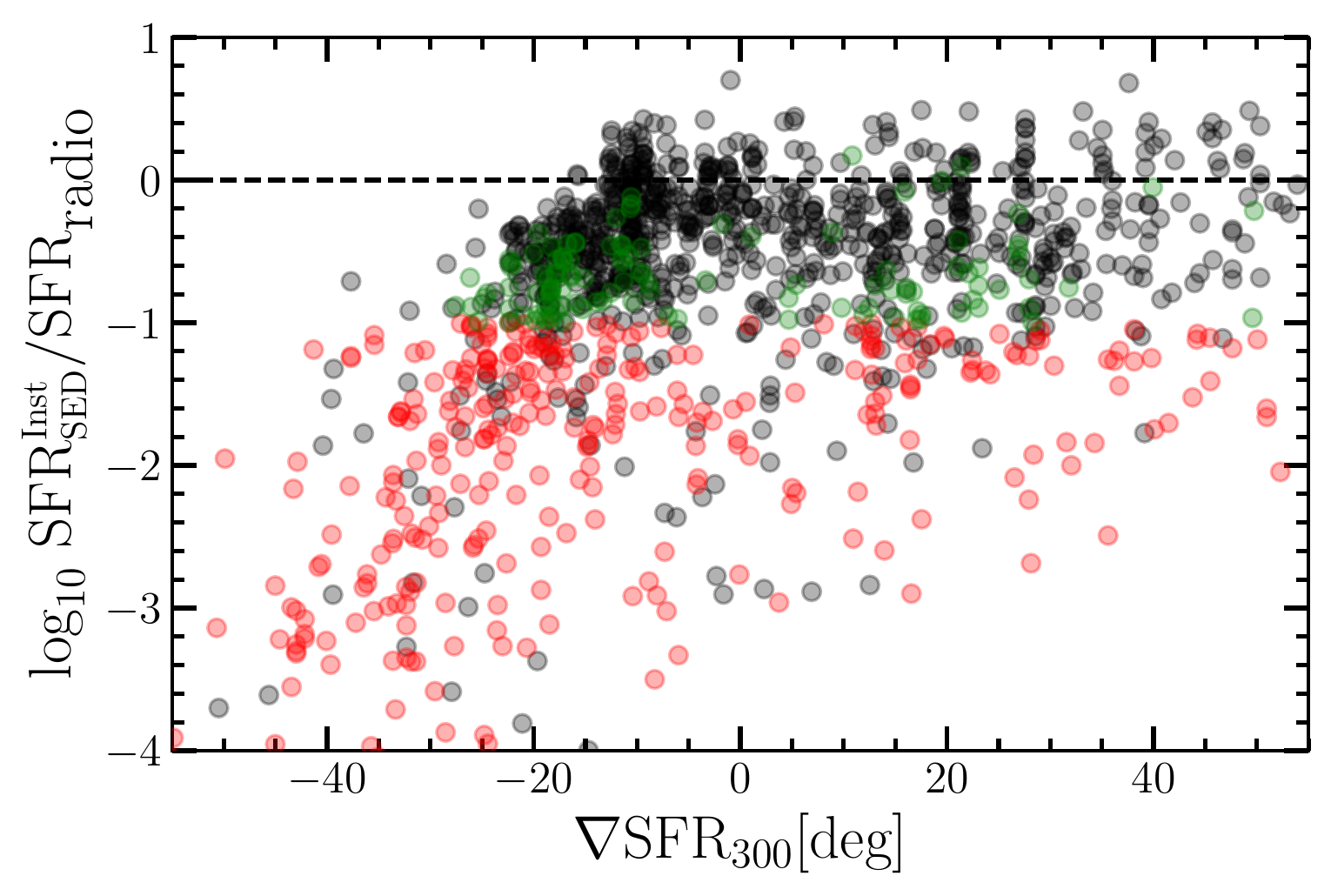}
    \caption{Ratio between the $\textrm{SFR}_{\textrm{SED}}^{\textrm{Inst}}$ and the SFR$_{\textrm{radio}}$ in function of the SFR gradient over the last 300\,Myr ($\nabla \textrm{SFR}_{300}$).  Black dots represent galaxies that agree in both SFRs indicators while green dots counts for outliers on the SFH delayed scenario and red dots counts for outliers for both SFH models. 90\% of the outliers would be selected as outliers either with flexible or non-parametric SFHs.}
    \label{outliers_check}
\end{figure}

As in the previous section, we find a population of galaxies with $\SFRradio{}$ larger than the one obtained here with the SED fitting, $\textrm{SFR}_{\textrm{SED}}^{\textrm{flex}}$.
We define the outliers galaxies such as the ratio $\textrm{log}_{10}~\textrm{SFR}_{\textrm{SED}}^{\textrm{flex}}/\textrm{SFR}_{\textrm{radio}}<-1$. We want to check if the galaxies with divergent SFR tracers are the same using the SFH delayed scenario or the non-parametric SFH one. The result is shown in Fig. \ref{outliers_check} with the green dots showing outliers in the SFH delayed scenario, while the red dots show outliers in both SFHs scenarios\footnote{We can not produce a figure similar to Fig.~\ref{outliers_check} using directly SFR$_{\textrm{SED}}^{\textrm{flex}}$ rather than SFR$_{\textrm{SED}}$ on the non-parametric SFH scenario, because the $\nabla \textrm{SFR}_{\Delta t}$ parameter is not available in output of \cigale\ for flexible SFH.}. A large fraction (90$\%$) of galaxies exhibit a very low $\textrm{SFR}_{\textrm{SED}}^{\rm inst}/\textrm{SFR}_{\textrm{radio}}$ ratio whether the parametric or non-parametric SFH is used to estimate the SED SFR. We conclude that the outliers remain the same whether we use analytical or non-parametric SFHs and our results are therefore independent from the choice of SFH model.

\subsection{Possible remaining AGN contamination? \label{AGN_model_ceck}}

\begin{figure}[h!]
  \begin{center}
    \includegraphics[width=\columnwidth]{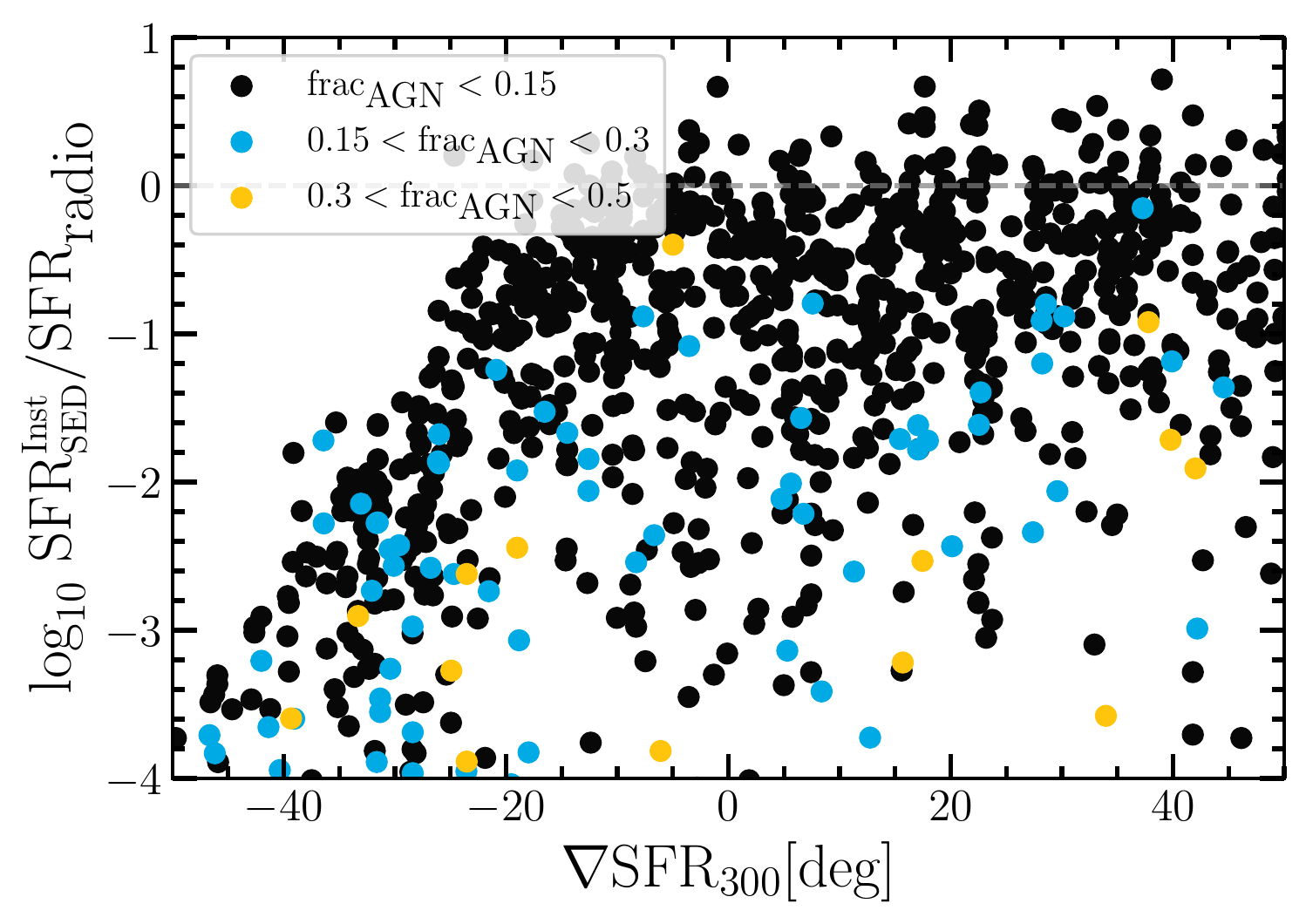}
  \end{center}
        \caption{Ratio between the SFR$_{\textrm{SED}}^{\textrm{Inst}}$ and the radio SFR as a function of the SFR gradient over the last 300\,Myr, as in Fig.~\ref{scatter_chi}. Color code represents the AGN fraction.}
        \label{scatter_AGN_fraction}
\end{figure}

As described in Sect.~\ref{AGN-selec}, \citet{Jimenez19} thoroughly applied a set of criteria to remove AGN host galaxies from their sample of pure star-forming galaxies. However, we add to this a last test fitting the galaxies using the AGN modeling module of \cigale, \texttt{skirtor} \citep{Stalevski16}. This \cigale\ module implements a modern clumpy two-phase torus model to compute the UV-to-IR SED model of AGN \citep[see][for further information]{Yang19,Buat21}.
From this SED modelling run, we collect the instantaneous SFR as well the AGN fraction (frac$_{\textrm{AGN}}$), which quantify the AGN contribution to the total IR luminosity (L$_{IR}$). Fig.~\ref{scatter_AGN_fraction} shows the ratios between $\textrm{SFR}_{\textrm{SED}}^{\textrm{Inst}}/\textrm{SFR}_{\textrm{radio}}$, color-coded with frac$_{\textrm{AGN}}$. 

As a results, $90\%$ of our galaxy sample have a $\textrm{frac}_{\textrm{AGN}}$ lower than 15\% (black dots) which is compatible with no AGN presence detected from the UV to IR multi-wavelength data \citep{Ciesla15}. Only $10\%$ of the galaxies have a value higher than 15\% (blue dots) and 2\% a frac$_{\rm AGN}$ larger than 30\% (yellow dots). We note that galaxies with weak to moderate AGN contribution ($>$15\%) have in general, ratios values lower than 0.1. However, even if we consider only the sources with frac$_{\textrm{AGN}}<15\%$, the results described in Sect.~\ref{res} are still valid, and we observe that SFR$_{\textrm{SED}}^{\textrm{Inst}}$ and SFR$_{\textrm{radio}}$ diverge in galaxies with decreasing SFH.

Despite having applied several criteria to remove radio sources with a significant AGN contribution (see Sect.~\ref{AGN-selec}), we can not rule out a remaining low level of AGN contamination in the radio emission. We note that further AGN classification would require a spectroscopic follow-up to establish additional criteria \citep[e.g.][]{BPT81,Best12}.

\subsection{Insights from other star-formation tracers}

We check the robustness of our results by adding two other star-formation tracers available in the COSMOS field, the IR luminosity and the H$_\alpha$ emission lines. Here, the IR luminosity is not an output of our \cigale\ run but comes from an independent IR SED fits of COSMOS galaxies. These two additional tracers are sensitive over different timescales.

\begin{figure}[ht]
    \centering
    \includegraphics[width=\columnwidth]{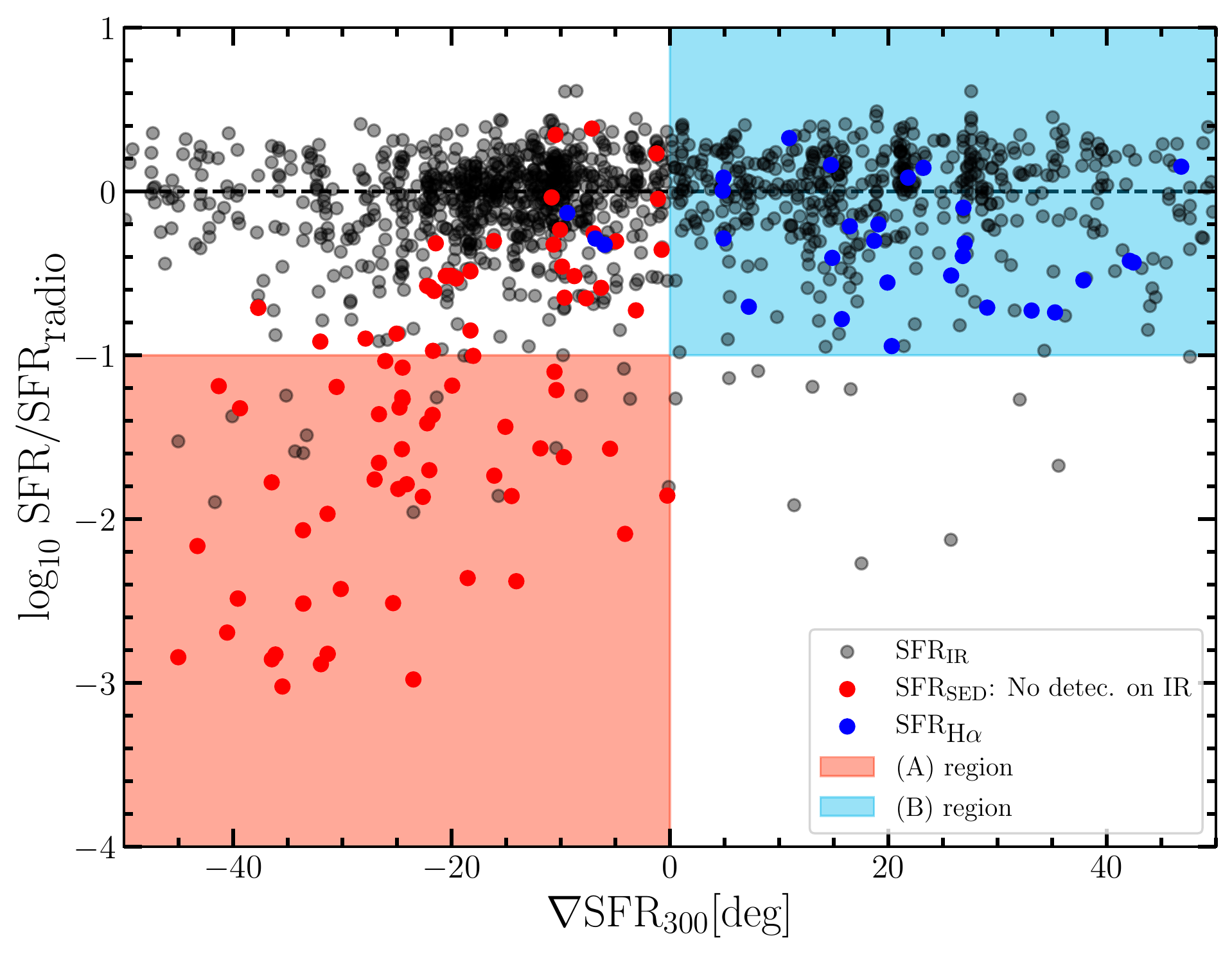}
    \caption{ Ratio between the SFR$_{\textrm{radio}}$ and $\textrm{SFR}$ from other tracers as a function of the SFR gradient over the last 300\,Myr ($\nabla \textrm{SFR}_{300}$). We use tree different SFR indicators: SFR$_{\textrm{IR}}$ with black circles, SFR$_{\textrm{H$_\alpha$}}$ with blue stars, and SFR$_{\textrm{SED}}^{\textrm{Inst}}$ with red circles. These galaxies marked with red circles are not detected in Herschel.}
    \label{otherTracers}
\end{figure}

As presented in Sect.~\ref{res}, the SFRs derived from radio are orders of magnitude larger than the ones derived from SED-fitting for galaxies with a declining SFH. One possible interpretation is that the radio traces the SFR over a longer timescale ($>100$\,Myr), while SED-fitting is considered to provide an instantaneous SFR. Given that H$_\alpha$ traces the SFR over short timescales \citep[$<10$\,Myr,][]{kennicutt12}, the SFR derived from H$_\alpha$ should follow more closely the instantaneous SFR from SED fitting. The spectroscopic compilation established by \citet{Saito20} includes public H$_\alpha$ emission line measurements, mainly from 3D-HST \citep[][]{Momcheva:2016lr}, and the zCOSMOS catalog of emission lines presented in \citet[][]{Silverman:2009} but extended to the final zCOSMOS-Bright spectroscopic sample. As in \citet{Saito20}\label, emission lines were corrected from dust attenuation derived using the SED fitting. Then, the H$_\alpha$ intrinsic luminosity was converted into SFR using \cite{kennicutt12}. Given the small size of the spectroscopic sample with H$_\alpha$ measured, we find only 29 matches to the radio sources. We find a good agreement between the SFR derived from H$_\alpha$ and radio (see blue dots in Fig.~\ref{otherTracers}). However, we find no source detected in H$_\alpha$ within the region showing a declining SFH. This result is consistent with low SFR values in the $\gradient<0$ part of the sample, as expected from SED-fitting. However, a robust conclusion would require a larger spectroscopic sample.

Another interpretation of the results discussed in Sect.~\ref{res} is that the $q_{IR}$ parameter used in Sect.~\ref{radiosfr} to convert the radio luminosity into SFR depends on the stellar mass or sSFR \citep[]{Gurkan18}. \citet[][]{Delvecchio21} find that $q_{IR}$ decreases with the stellar mass. However, we estimate that this would not explain more than 0.3\,dex overestimate for the $\SFRradio{}$ for massive galaxies, which is not sufficient to correct the difference seen at low $\gradient{}$. We can directly test the validity of the $q_{IR}$ factor with our sample. We use the super-deblended FIR catalog from \citet{Jin18} including MIPS \citep[]{lefloch09}, as well as Herschel data from the PEP survey \citep[]{Lutz2011} at 100 and 160\,$\mu$m, and the HerMES survey \citep[]{Oliver12} at 250, 350, and 500\,$\mu$m. The fit over the full wavelength range is performed  with \texttt{Le Phare} \citep{Arnouts02, Ilbert06} exactly in the same way as described in \citet{Ilbert15} using a combination of \citet[]{BruzualCharlot03} and \citet[]{Dale02} templates. The infrared luminosity is based on the integral of the SED from 8  to 1000\,$\mu$m. Fig.~\ref{otherTracers} shows the comparison between SFR$_{\textrm{IR}}$ and SFR$_{\rm{radio}}$. We find an excellent agreement, even in the regime of declining SFH. Given that the far-IR is sensitive over a long timescale \citep[$\sim100$ millions years,][]{kennicutt12}, as a consequence of old stars heating the dust, we conclude that such agreement between radio and IR tracers is expected. Moreover, we find that the population without any Herschel counterpart is well located within the region with a declining SFH, as expected. The existence of this population gives a hint that the radio tracer could be sensitive over longer timescale than the far-IR one for a significant population of galaxies.

\section{Conclusions and perspectives}

We used a mass-complete sub-sample of star-forming galaxies drawn from the VLA 3\,GHz project, and described in \citet{Jimenez19}, to investigate the difference between the SFR derived from radio observations and the one derived from UV-to-IR SED modeling. We used the SED modeling code \cigale\ that includes the new non-parametric SFH models \citep{Ciesla22}. 

The results of our analysis can be summarized as follow:
\begin{itemize}
    \item Approximately 30\% our galaxies present a radio SFR ten times larger than the instantaneous SED SFR. This trend affects primarily the galaxies that show a declining SFH activity over the last 300\,Myr, i.e. on their way to quench. 
    \item By averaging the SFH over different timescales, we found that both SFR indicators converge toward a consistent value, when the SFHs obtained from the best fits are averaged over a period larger than 150\,Myr to derive SFR$_{\textrm{SED}}$.
    \item From a set of tests, we show that these results are independent from the choice of SFH model, and are not due to persistent AGN contamination.
    \item For the sources detected with Herschel, we find a good agreement between the SFR derived from far-infrared and radio, even for declining SFH. We conclude that the  $q_{IR}$ dependency with stellar mass does not explain our result and that 1.4\,GHz traces star-formation over similar or longer timescale than infrared.
\end{itemize}

These results and in particular the discrepancies between the SFRs indicators suggest that the SFRs obtained via radio observations at 1.4\,GHz rest-frame, trace a longer timescale with respect to the SFR derive from the SED-template fitting, considered here as an instantaneous SFR indicator. As presented in Fig. \ref{ratios_distributions}, the SFH derived from SED-fitting needs to be integrated on period longer than 150-300\,Myr to be in agreement with the radio SFR. On shorter timescales, the discrepancy between the two estimators affects primarily galaxies with a declining SFH. This can be problematic for measuring SFR in rapidly quenched galaxies. 
On the other hand, this divergence of SFR$_{\textrm{SED}}^{\textrm{Inst}}$ and $\SFRradio{}$ for a sample free from AGN contamination could be a good criteria to isolate post-starburst galaxies.

Two mechanisms could explain the different timescale between the radio tracer and the SFR derived via SED-template fitting. Firstly, there is a time delay between the beginning of star-formation and supernova explosions, up to 30\,Myr, assuming stars more massive than 8\,$\Msol$ explode as SNe. So, the radio emission is necessarily lagging behind emission in UV or IR. Secondly, the radio emission is due to the synchrotron emission of CRe accelerated by SN and trapped in magnetic fields \citep[e.g.,][]{Murphy09}. Therefore, such radio emission could occur over long timescales, related to the timescale during which CRe travel through the ISM. The cooling timescale due to synchrotron for CRe emitting at 1 GHz, could be of the order 30-100\,Myr \citep{Condon92, Murphy09}. These two factors could explain that the radio wavelength still trace the star-formation up to 100\,Myr after a quenching. Still, our results show that such timescale could be over 150\,Myr for the specific population with low sSFR, possibly undergoing a quenching.

On the other hand, the turbulence of the galactic magnetic fields can be trigger by gravitational instabilities during mergers \citep[e.g.,][]{Drzazga_2011, Rajpurohit_2021}, can generate an excess of synchrotron radiation compared to undisturbed magnetic fields, this via scattering process and confinement of CRe \citep{Lisenfeld10,Heesen14,Reichherzer_2021}. A morphological study of our galaxy sample could help to understand the discrepancy between the SFRs indicators studied in this work. One of the most favored candidates to develop this perspective would be the imagery data from the \textit{James Webb} Space Telescope \citep[JWST,][]{Gardner_2006, Snyder_2019, Sailor21}. 

We conclude that the use 1.4\,GHz as star-formation tracer still need some investigation, specially for galaxy populations with a declining SFH. A better understanding on how the radio wavelength traces the star-formation is becoming crucial, with The Square Kilometre Array 
(SKA)\footnote{\url{https://www.skao.int/en/science-users}} being able to measure and study the radio continuum of millions of radio sources, even in the very deep universe \citep{Murphy09,Jarvis15}.

\begin{acknowledgements}
We thank the referee for his/her comments which helped improve the paper.
We also thank Alessandro Boselli, David Elbaz, and Vernesa Smol\v ci\'c for insightful ideas and discussions.
This project has received financial support from the CNRS through the MITI interdisciplinary programs.
We warmly acknowledge the contributions of the entire COSMOS collaboration consisting of more than 100 scientists. The HST-COSMOS program was supported through NASA grant HST-GO-09822. More information on the COSMOS survey is available at \url{https://cosmos.astro.caltech.edu}. This research is also partly supported by the Centre National d'Etudes Spatiales (CNES). OI acknowledges the funding of the French Agence Nationale de la Recherche for the project iMAGE (grant ANR-22-CE31-0007). 

\end{acknowledgements}

\bibliographystyle{aa_url}
\bibliography{aanda}

\begin{thebibliography}{96}
\expandafter\ifx\csname natexlab\endcsname\relax\def\natexlab#1{#1}\fi

\bibitem[{{Arnouts} {et~al.}(2013){Arnouts}, {Le Floc'h}, {Chevallard},
  {Johnson}, {Ilbert}, {Treyer}, {Aussel}, {Capak}, {Sanders}, {Scoville},
  {McCracken}, {Milliard}, {Pozzetti}, \& {Salvato}}]{Arnouts13}
{Arnouts}, S., {Le Floc'h}, E., {Chevallard}, J., {et~al.} 2013,
  \href{http://dx.doi.org/10.1051/0004-6361/201321768}{\color{magenta}\aap},
  \href{https://ui.adsabs.harvard.edu/abs/2013A&A...558A..67A}{558, A67}

\bibitem[{{Arnouts} {et~al.}(2002){Arnouts}, {Moscardini}, {Vanzella},
  {Colombi}, {Cristiani}, {Fontana}, {Giallongo}, {Matarrese}, \&
  {Saracco}}]{Arnouts02}
{Arnouts}, S., {Moscardini}, L., {Vanzella}, E., {et~al.} 2002,
  \href{http://dx.doi.org/10.1046/j.1365-8711.2002.04988.x}{\color{magenta}\mnras},
  \href{https://ui.adsabs.harvard.edu/abs/2002MNRAS.329..355A}{329, 355}

\bibitem[{{Aufort} {et~al.}(2020){Aufort}, {Ciesla}, {Pudlo}, \&
  {Buat}}]{Aufort20}
{Aufort}, G., {Ciesla}, L., {Pudlo}, P., \& {Buat}, V. 2020,
  \href{http://dx.doi.org/10.1051/0004-6361/201936788}{\color{magenta}\aap},
  \href{https://ui.adsabs.harvard.edu/abs/2020A&A...635A.136A}{635, A136}

\bibitem[{{Baldwin} {et~al.}(1981){Baldwin}, {Phillips}, \&
  {Terlevich}}]{BPT81}
{Baldwin}, J.~A., {Phillips}, M.~M., \& {Terlevich}, R. 1981,
  \href{http://dx.doi.org/10.1086/130766}{\color{magenta}\pasp},
  \href{https://ui.adsabs.harvard.edu/abs/1981PASP...93....5B}{93, 5}

\bibitem[{{Bell} {et~al.}(2005){Bell}, {Papovich}, {Wolf}, {Le Floc'h},
  {Caldwell}, {Barden}, {Egami}, {McIntosh}, {Meisenheimer},
  {P{\'e}rez-Gonz{\'a}lez}, {Rieke}, {Rieke}, {Rigby}, \& {Rix}}]{Bell05}
{Bell}, E.~F., {Papovich}, C., {Wolf}, C., {et~al.} 2005,
  \href{http://dx.doi.org/10.1086/429552}{\color{magenta}\apj},
  \href{https://ui.adsabs.harvard.edu/abs/2005ApJ...625...23B}{625, 23}

\bibitem[{{Berta} {et~al.}(2013){Berta}, {Lutz}, {Santini}, {Wuyts}, {Rosario},
  {Brisbin}, {Cooray}, {Franceschini}, {Gruppioni}, {Hatziminaoglou}, {Hwang},
  {Le Floc'h}, {Magnelli}, {Nordon}, {Oliver}, {Page}, {Popesso}, {Pozzetti},
  {Pozzi}, {Riguccini}, {Rodighiero}, {Roseboom}, {Scott}, {Symeonidis},
  {Valtchanov}, {Viero}, \& {Wang}}]{Berta13}
{Berta}, S., {Lutz}, D., {Santini}, P., {et~al.} 2013,
  \href{http://dx.doi.org/10.1051/0004-6361/201220859}{\color{magenta}\aap},
  \href{https://ui.adsabs.harvard.edu/abs/2013A&A...551A.100B}{551, A100}

\bibitem[{{Best} \& {Heckman}(2012)}]{Best12}
{Best}, P.~N. \& {Heckman}, T.~M. 2012,
  \href{http://dx.doi.org/10.1111/j.1365-2966.2012.20414.x}{\color{magenta}\mnras},
  \href{https://ui.adsabs.harvard.edu/abs/2012MNRAS.421.1569B}{421, 1569}

\bibitem[{{Boissier} {et~al.}(2007){Boissier}, {Gil de Paz}, {Boselli},
  {Madore}, {Buat}, {Cortese}, {Burgarella}, {Mu{\~n}oz-Mateos}, {Barlow},
  {Forster}, {Friedman}, {Martin}, {Morrissey}, {Neff}, {Schiminovich},
  {Seibert}, {Small}, {Wyder}, {Bianchi}, {Donas}, {Heckman}, {Lee},
  {Milliard}, {Rich}, {Szalay}, {Welsh}, \& {Yi}}]{Boissier07}
{Boissier}, S., {Gil de Paz}, A., {Boselli}, A., {et~al.} 2007,
  \href{http://dx.doi.org/10.1086/516642}{\color{magenta}\apjs},
  \href{https://ui.adsabs.harvard.edu/abs/2007ApJS..173..524B}{173, 524}

\bibitem[{{Bolzonella} {et~al.}(2000){Bolzonella}, {Miralles}, \&
  {Pell{\'o}}}]{bolzonella00}
{Bolzonella}, M., {Miralles}, J.~M., \& {Pell{\'o}}, R. 2000, \aap,
  \href{https://ui.adsabs.harvard.edu/abs/2000A&A...363..476B}{363, 476}

\bibitem[{{Boquien} {et~al.}(2014){Boquien}, {Buat}, \& {Perret}}]{Boquien14}
{Boquien}, M., {Buat}, V., \& {Perret}, V. 2014,
  \href{http://dx.doi.org/10.1051/0004-6361/201424441}{\color{magenta}\aap},
  \href{https://ui.adsabs.harvard.edu/abs/2014A&A...571A..72B}{571, A72}

\bibitem[{{Boquien} {et~al.}(2019){Boquien}, {Burgarella}, {Roehlly}, {Buat},
  {Ciesla}, {Corre}, {Inoue}, \& {Salas}}]{Boquien19}
{Boquien}, M., {Burgarella}, D., {Roehlly}, Y., {et~al.} 2019,
  \href{http://dx.doi.org/10.1051/0004-6361/201834156}{\color{magenta}\aap},
  \href{https://ui.adsabs.harvard.edu/abs/2019A&A...622A.103B}{622, A103}

\bibitem[{{Bruzual} \& {Charlot}(2003)}]{BruzualCharlot03}
{Bruzual}, G. \& {Charlot}, S. 2003,
  \href{http://dx.doi.org/10.1046/j.1365-8711.2003.06897.x}{\color{magenta}\mnras},
  \href{https://ui.adsabs.harvard.edu/abs/2003MNRAS.344.1000B}{344, 1000}

\bibitem[{{Buat} {et~al.}(2019){Buat}, {Ciesla}, {Boquien}, {Ma{\l}ek}, \&
  {Burgarella}}]{Buat19}
{Buat}, V., {Ciesla}, L., {Boquien}, M., {Ma{\l}ek}, K., \& {Burgarella}, D.
  2019,
  \href{http://dx.doi.org/10.1051/0004-6361/201936643}{\color{magenta}\aap},
  \href{https://ui.adsabs.harvard.edu/abs/2019A&A...632A..79B}{632, A79}

\bibitem[{{Buat} {et~al.}(2014){Buat}, {Heinis}, {Boquien}, {Burgarella},
  {Charmandaris}, {Boissier}, {Boselli}, {Le Borgne}, \& {Morrison}}]{Buat14}
{Buat}, V., {Heinis}, S., {Boquien}, M., {et~al.} 2014,
  \href{http://dx.doi.org/10.1051/0004-6361/201322081}{\color{magenta}\aap},
  \href{https://ui.adsabs.harvard.edu/abs/2014A&A...561A..39B}{561, A39}

\bibitem[{{Buat} {et~al.}(2005){Buat}, {Iglesias-P{\'a}ramo}, {Seibert},
  {Burgarella}, {Charlot}, {Martin}, {Xu}, {Heckman}, {Boissier}, {Boselli},
  {Barlow}, {Bianchi}, {Byun}, {Donas}, {Forster}, {Friedman}, {Jelinski},
  {Lee}, {Madore}, {Malina}, {Milliard}, {Morissey}, {Neff}, {Rich},
  {Schiminovitch}, {Siegmund}, {Small}, {Szalay}, {Welsh}, \& {Wyder}}]{Buat05}
{Buat}, V., {Iglesias-P{\'a}ramo}, J., {Seibert}, M., {et~al.} 2005,
  \href{http://dx.doi.org/10.1086/423241}{\color{magenta}\apjl},
  \href{https://ui.adsabs.harvard.edu/abs/2005ApJ...619L..51B}{619, L51}

\bibitem[{{Buat} {et~al.}(2021){Buat}, {Mountrichas}, {Yang}, {Boquien},
  {Roehlly}, {Burgarella}, {Stalevski}, {Ciesla}, \& {Theul{\'e}}}]{Buat21}
{Buat}, V., {Mountrichas}, G., {Yang}, G., {et~al.} 2021,
  \href{http://dx.doi.org/10.1051/0004-6361/202141797}{\color{magenta}\aap},
  \href{https://ui.adsabs.harvard.edu/abs/2021A&A...654A..93B}{654, A93}

\bibitem[{{Calzetti} {et~al.}(2000){Calzetti}, {Armus}, {Bohlin}, {Kinney},
  {Koornneef}, \& {Storchi-Bergmann}}]{Calzetti00}
{Calzetti}, D., {Armus}, L., {Bohlin}, R.~C., {et~al.} 2000,
  \href{http://dx.doi.org/10.1086/308692}{\color{magenta}\apj},
  \href{https://ui.adsabs.harvard.edu/abs/2000ApJ...533..682C}{533, 682}

\bibitem[{{Chabrier}(2003)}]{Chabrier03}
{Chabrier}, G. 2003,
  \href{http://dx.doi.org/10.1086/376392}{\color{magenta}\pasp},
  \href{https://ui.adsabs.harvard.edu/abs/2003PASP..115..763C}{115, 763}

\bibitem[{{Chevallard} \& {Charlot}(2016)}]{Chevallard16}
{Chevallard}, J. \& {Charlot}, S. 2016,
  \href{http://dx.doi.org/10.1093/mnras/stw1756}{\color{magenta}\mnras},
  \href{https://ui.adsabs.harvard.edu/abs/2016MNRAS.462.1415C}{462, 1415}

\bibitem[{{Ciesla} {et~al.}(2021){Ciesla}, {Buat}, {Boquien}, {Boselli},
  {Elbaz}, \& {Aufort}}]{Ciesla21}
{Ciesla}, L., {Buat}, V., {Boquien}, M., {et~al.} 2021,
  \href{http://dx.doi.org/10.1051/0004-6361/202140762}{\color{magenta}\aap},
  \href{https://ui.adsabs.harvard.edu/abs/2021A&A...653A...6C}{653, A6}

\bibitem[{{Ciesla} {et~al.}(2015){Ciesla}, {Charmandaris}, {Georgakakis},
  {Bernhard}, {Mitchell}, {Buat}, {Elbaz}, {LeFloc'h}, {Lacey}, {Magdis}, \&
  {Xilouris}}]{Ciesla15}
{Ciesla}, L., {Charmandaris}, V., {Georgakakis}, A., {et~al.} 2015,
  \href{http://dx.doi.org/10.1051/0004-6361/201425252}{\color{magenta}\aap},
  \href{https://ui.adsabs.harvard.edu/abs/2015A&A...576A..10C}{576, A10}

\bibitem[{{Ciesla} {et~al.}(2017){Ciesla}, {Elbaz}, \& {Fensch}}]{Ciesla17}
{Ciesla}, L., {Elbaz}, D., \& {Fensch}, J. 2017,
  \href{http://dx.doi.org/10.1051/0004-6361/201731036}{\color{magenta}\aap},
  \href{https://ui.adsabs.harvard.edu/abs/2017A&A...608A..41C}{608, A41}

\bibitem[{{Ciesla} {et~al.}(2018){Ciesla}, {Elbaz}, {Schreiber}, {Daddi}, \&
  {Wang}}]{Ciesla18}
{Ciesla}, L., {Elbaz}, D., {Schreiber}, C., {Daddi}, E., \& {Wang}, T. 2018,
  \href{http://dx.doi.org/10.1051/0004-6361/201832715}{\color{magenta}\aap},
  \href{https://ui.adsabs.harvard.edu/abs/2018A&A...615A..61C}{615, A61}

\bibitem[{{Ciesla} {et~al.}(2022){Ciesla}, {G{\'o}mez-Guijarro}, {Buat},
  {Elbaz}, {Jin}, {B{\'e}thermin}, {Daddi}, {Franco}, {Inami}, {Magdis}, \&
  {Magnelli}}]{Ciesla22}
{Ciesla}, L., {G{\'o}mez-Guijarro}, C., {Buat}, V., {et~al.} 2022,
  \href{https://ui.adsabs.harvard.edu/abs/2022arXiv221102510C}{arXiv e-prints,
  arXiv:2211.02510}

\bibitem[{{Condon}(1992)}]{Condon92}
{Condon}, J.~J. 1992,
  \href{http://dx.doi.org/10.1146/annurev.aa.30.090192.003043}{\color{magenta}\araa},
  \href{https://ui.adsabs.harvard.edu/abs/1992ARA&A..30..575C}{30, 575}

\bibitem[{{Condon} {et~al.}(1991){Condon}, {Anderson}, \& {Helou}}]{Condon91}
{Condon}, J.~J., {Anderson}, M.~L., \& {Helou}, G. 1991,
  \href{http://dx.doi.org/10.1086/170258}{\color{magenta}\apj},
  \href{https://ui.adsabs.harvard.edu/abs/1991ApJ...376...95C}{376, 95}

\bibitem[{{da Cunha} {et~al.}(2008){da Cunha}, {Charlot}, \&
  {Elbaz}}]{dacunha08}
{da Cunha}, E., {Charlot}, S., \& {Elbaz}, D. 2008,
  \href{http://dx.doi.org/10.1111/j.1365-2966.2008.13535.x}{\color{magenta}\mnras},
  \href{https://ui.adsabs.harvard.edu/abs/2008MNRAS.388.1595D}{388, 1595}

\bibitem[{{Dale} \& {Helou}(2002)}]{Dale02}
{Dale}, D.~A. \& {Helou}, G. 2002,
  \href{http://dx.doi.org/10.1086/341632}{\color{magenta}\apj},
  \href{https://ui.adsabs.harvard.edu/abs/2002ApJ...576..159D}{576, 159}

\bibitem[{{Dale} {et~al.}(2014){Dale}, {Helou}, {Magdis}, {Armus},
  {D{\'\i}az-Santos}, \& {Shi}}]{Dale14}
{Dale}, D.~A., {Helou}, G., {Magdis}, G.~E., {et~al.} 2014,
  \href{http://dx.doi.org/10.1088/0004-637X/784/1/83}{\color{magenta}\apj},
  \href{https://ui.adsabs.harvard.edu/abs/2014ApJ...784...83D}{784, 83}

\bibitem[{{Delhaize} {et~al.}(2017){Delhaize}, {Smol{\v{c}}i{\'c}},
  {Delvecchio}, {Novak}, {Sargent}, {Baran}, {Magnelli}, {Zamorani},
  {Schinnerer}, {Murphy}, {Aravena}, {Berta}, {Bondi}, {Capak}, {Carilli},
  {Ciliegi}, {Civano}, {Ilbert}, {Karim}, {Laigle}, {Le F{\`e}vre}, {Marchesi},
  {McCracken}, {Salvato}, {Seymour}, \& {Tasca}}]{Delhaize17}
{Delhaize}, J., {Smol{\v{c}}i{\'c}}, V., {Delvecchio}, I., {et~al.} 2017,
  \href{http://dx.doi.org/10.1051/0004-6361/201629430}{\color{magenta}\aap},
  \href{https://ui.adsabs.harvard.edu/abs/2017A&A...602A...4D}{602, A4}

\bibitem[{{Delvecchio} {et~al.}(2021){Delvecchio}, {Daddi}, {Sargent},
  {Jarvis}, {Elbaz}, {Jin}, {Liu}, {Whittam}, {Algera}, {Carraro}, {D'Eugenio},
  {Delhaize}, {Kalita}, {Leslie}, {Moln{\'a}r}, {Novak}, {Prandoni},
  {Smol{\v{c}}i{\'c}}, {Ao}, {Aravena}, {Bournaud}, {Collier},
  {Randriamampandry}, {Randriamanakoto}, {Rodighiero}, {Schober}, {White}, \&
  {Zamorani}}]{Delvecchio21}
{Delvecchio}, I., {Daddi}, E., {Sargent}, M.~T., {et~al.} 2021,
  \href{http://dx.doi.org/10.1051/0004-6361/202039647}{\color{magenta}\aap},
  \href{https://ui.adsabs.harvard.edu/abs/2021A&A...647A.123D}{647, A123}

\bibitem[{{Delvecchio} {et~al.}(2014){Delvecchio}, {Gruppioni}, {Pozzi},
  {Berta}, {Zamorani}, {Cimatti}, {Lutz}, {Scott}, {Vignali}, {Cresci},
  {Feltre}, {Cooray}, {Vaccari}, {Fritz}, {Le Floc'h}, {Magnelli}, {Popesso},
  {Oliver}, {Bock}, {Carollo}, {Contini}, {Le F{\'e}vre}, {Lilly}, {Mainieri},
  {Renzini}, \& {Scodeggio}}]{Delvecchio14}
{Delvecchio}, I., {Gruppioni}, C., {Pozzi}, F., {et~al.} 2014,
  \href{http://dx.doi.org/10.1093/mnras/stu130}{\color{magenta}\mnras},
  \href{https://ui.adsabs.harvard.edu/abs/2014MNRAS.439.2736D}{439, 2736}

\bibitem[{{Delvecchio} {et~al.}(2017){Delvecchio}, {Smol{\v{c}}i{\'c}},
  {Zamorani}, {Lagos}, {Berta}, {Delhaize}, {Baran}, {Alexander}, {Rosario},
  {Gonzalez-Perez}, {Ilbert}, {Lacey}, {Le F{\`e}vre}, {Miettinen}, {Aravena},
  {Bondi}, {Carilli}, {Ciliegi}, {Mooley}, {Novak}, {Schinnerer}, {Capak},
  {Civano}, {Fanidakis}, {Herrera Ruiz}, {Karim}, {Laigle}, {Marchesi},
  {McCracken}, {Middleberg}, {Salvato}, \& {Tasca}}]{Delvecchio17}
{Delvecchio}, I., {Smol{\v{c}}i{\'c}}, V., {Zamorani}, G., {et~al.} 2017,
  \href{http://dx.doi.org/10.1051/0004-6361/201629367}{\color{magenta}\aap},
  \href{https://ui.adsabs.harvard.edu/abs/2017A&A...602A...3D}{602, A3}

\bibitem[{{Dom{\'\i}nguez S{\'a}nchez} {et~al.}(2012){Dom{\'\i}nguez
  S{\'a}nchez}, {Mignoli}, {Pozzi}, {Calura}, {Cimatti}, {Gruppioni}, {Cepa},
  {S{\'a}nchez Portal}, {Zamorani}, {Berta}, {Elbaz}, {Le Floc'h}, {Granato},
  {Lutz}, {Maiolino}, {Matteucci}, {Nair}, {Nordon}, {Pozzetti}, {Silva},
  {Silverman}, {Wuyts}, {Carollo}, {Contini}, {Kneib}, {Le F{\`e}vre}, {Lilly},
  {Mainieri}, {Renzini}, {Scodeggio}, {Bardelli}, {Bolzonella}, {Bongiorno},
  {Caputi}, {Coppa}, {Cucciati}, {de la Torre}, {de Ravel}, {Franzetti},
  {Garilli}, {Iovino}, {Kampczyk}, {Knobel}, {Kova{\v{c}}}, {Lamareille}, {Le
  Borgne}, {Le Brun}, {Maier}, {Magnelli}, {Pell{\'o}}, {Peng},
  {Perez-Montero}, {Ricciardelli}, {Riguccini}, {Tanaka}, {Tasca}, {Tresse},
  {Vergani}, \& {Zucca}}]{DominguezSanchez12}
{Dom{\'\i}nguez S{\'a}nchez}, H., {Mignoli}, M., {Pozzi}, F., {et~al.} 2012,
  \href{http://dx.doi.org/10.1111/j.1365-2966.2012.21710.x}{\color{magenta}\mnras},
  \href{https://ui.adsabs.harvard.edu/abs/2012MNRAS.426..330D}{426, 330}

\bibitem[{{Donley} {et~al.}(2012){Donley}, {Koekemoer}, {Brusa}, {Capak},
  {Cardamone}, {Civano}, {Ilbert}, {Impey}, {Kartaltepe}, {Miyaji}, {Salvato},
  {Sanders}, {Trump}, \& {Zamorani}}]{Donley12}
{Donley}, J.~L., {Koekemoer}, A.~M., {Brusa}, M., {et~al.} 2012,
  \href{http://dx.doi.org/10.1088/0004-637X/748/2/142}{\color{magenta}\apj},
  \href{https://ui.adsabs.harvard.edu/abs/2012ApJ...748..142D}{748, 142}

\bibitem[{{Drzazga} {et~al.}(2011){Drzazga}, {Chy{\.z}y}, {Jurusik}, \&
  {Wi{\'o}rkiewicz}}]{Drzazga_2011}
{Drzazga}, R.~T., {Chy{\.z}y}, K.~T., {Jurusik}, W., \& {Wi{\'o}rkiewicz}, K.
  2011,
  \href{http://dx.doi.org/10.1051/0004-6361/201016092}{\color{magenta}\aap},
  \href{https://ui.adsabs.harvard.edu/abs/2011A&A...533A..22D}{533, A22}

\bibitem[{{Figueira} {et~al.}(2022){Figueira}, {Pollo}, {Ma{\l}ek}, {Buat},
  {Boquien}, {Pistis}, {Cassar{\`a}}, {Vergani}, {Hamed}, \&
  {Salim}}]{Figueira22}
{Figueira}, M., {Pollo}, A., {Ma{\l}ek}, K., {et~al.} 2022,
  \href{http://dx.doi.org/10.1051/0004-6361/202141701}{\color{magenta}\aap},
  \href{https://ui.adsabs.harvard.edu/abs/2022A&A...667A..29F}{667, A29}

\bibitem[{{Finkelstein} {et~al.}(2022){Finkelstein}, {Bagley}, {Haro},
  {Dickinson}, {Ferguson}, {Kartaltepe}, {Papovich}, {Burgarella}, {Kocevski},
  {Huertas-Company}, {Iyer}, {Koekemoer}, {Larson}, {P{\'e}rez-Gonz{\'a}lez},
  {Rose}, {Tacchella}, {Wilkins}, {Chworowsky}, {Medrano}, {Morales},
  {Somerville}, {Yung}, {Fontana}, {Giavalisco}, {Grazian}, {Grogin}, {Kewley},
  {Kirkpatrick}, {Kurczynski}, {Lotz}, {Pentericci}, {Pirzkal}, {Ravindranath},
  {Ryan}, {Trump}, {Yang}, {Almaini}, {Amor{\'\i}n}, {Annunziatella},
  {Backhaus}, {Barro}, {Behroozi}, {Bell}, {Bhatawdekar}, {Bisigello}, {Bromm},
  {Buat}, {Buitrago}, {Calabr{\`o}}, {Casey}, {Castellano}, {Ch{\'a}vez Ortiz},
  {Ciesla}, {Cleri}, {Cohen}, {Cole}, {Cooke}, {Cooper}, {Cooray}, {Costantin},
  {Cox}, {Croton}, {Daddi}, {Dav{\'e}}, {de La Vega}, {Dekel}, {Elbaz},
  {Estrada-Carpenter}, {Faber}, {Fern{\'a}ndez}, {Finkelstein}, {Freundlich},
  {Fujimoto}, {Garc{\'\i}a-Argum{\'a}nez}, {Gardner}, {Gawiser},
  {G{\'o}mez-Guijarro}, {Guo}, {Hamblin}, {Hamilton}, {Hathi}, {Holwerda},
  {Hirschmann}, {Hutchison}, {Jaskot}, {Jha}, {Jogee}, {Juneau}, {Jung},
  {Kassin}, {Le Bail}, {Leung}, {Lucas}, {Magnelli}, {Mantha}, {Matharu},
  {McGrath}, {McIntosh}, {Merlin}, {Mobasher}, {Newman}, {Nicholls}, {Pandya},
  {Rafelski}, {Ronayne}, {Santini}, {Seill{\'e}}, {Shah}, {Shen}, {Simons},
  {Snyder}, {Stanway}, {Straughn}, {Teplitz}, {Vanderhoof}, {Vega-Ferrero},
  {Wang}, {Weiner}, {Willmer}, {Wuyts}, {Zavala}, \& {CEERS
  Team}}]{Finkelstein22}
{Finkelstein}, S.~L., {Bagley}, M.~B., {Haro}, P.~A., {et~al.} 2022,
  \href{http://dx.doi.org/10.3847/2041-8213/ac966e}{\color{magenta}\apjl},
  \href{https://ui.adsabs.harvard.edu/abs/2022ApJ...940L..55F}{940, L55}

\bibitem[{Gardner {et~al.}(2006)Gardner, Mather, Clampin, Doyon, Greenhouse,
  Hammel, Hutchings, Jakobsen, Lilly, Long, Lunine, Mccaughrean, Mountain,
  Nella, Rieke, Rieke, Rix, Smith, Sonneborn, Stiavelli, Stockman, Windhorst,
  \& Wright}]{Gardner_2006}
Gardner, J.~P., Mather, J.~C., Clampin, M., {et~al.} 2006,
  \href{http://dx.doi.org/10.1007/s11214-006-8315-7}{\color{magenta}Space
  Science Reviews}, 123, 485

\bibitem[{{G{\"u}rkan} {et~al.}(2015){G{\"u}rkan}, {Hardcastle}, {Jarvis},
  {Smith}, {Bourne}, {Dunne}, {Maddox}, {Ivison}, \& {Fritz}}]{Gurkan15}
{G{\"u}rkan}, G., {Hardcastle}, M.~J., {Jarvis}, M.~J., {et~al.} 2015,
  \href{http://dx.doi.org/10.1093/mnras/stv1502}{\color{magenta}\mnras},
  \href{https://ui.adsabs.harvard.edu/abs/2015MNRAS.452.3776G}{452, 3776}

\bibitem[{{G{\"u}rkan} {et~al.}(2018){G{\"u}rkan}, {Hardcastle}, {Smith},
  {Best}, {Bourne}, {Calistro-Rivera}, {Heald}, {Jarvis}, {Prandoni},
  {R{\"o}ttgering}, {Sabater}, {Shimwell}, {Tasse}, \& {Williams}}]{Gurkan18}
{G{\"u}rkan}, G., {Hardcastle}, M.~J., {Smith}, D.~J.~B., {et~al.} 2018,
  \href{http://dx.doi.org/10.1093/mnras/sty016}{\color{magenta}\mnras},
  \href{https://ui.adsabs.harvard.edu/abs/2018MNRAS.475.3010G}{475, 3010}

\bibitem[{{Heesen} {et~al.}(2014){Heesen}, {Brinks}, {Leroy}, {Heald}, {Braun},
  {Bigiel}, \& {Beck}}]{Heesen14}
{Heesen}, V., {Brinks}, E., {Leroy}, A.~K., {et~al.} 2014,
  \href{http://dx.doi.org/10.1088/0004-6256/147/5/103}{\color{magenta}\aj},
  \href{https://ui.adsabs.harvard.edu/abs/2014AJ....147..103H}{147, 103}

\bibitem[{{Helou} {et~al.}(1985){Helou}, {Soifer}, \&
  {Rowan-Robinson}}]{Helou85}
{Helou}, G., {Soifer}, B.~T., \& {Rowan-Robinson}, M. 1985,
  \href{http://dx.doi.org/10.1086/184556}{\color{magenta}\apjl},
  \href{https://ui.adsabs.harvard.edu/abs/1985ApJ...298L...7H}{298, L7}

\bibitem[{{Ilbert} {et~al.}(2015){Ilbert}, {Arnouts}, {Le Floc'h}, {Aussel},
  {Bethermin}, {Capak}, {Hsieh}, {Kajisawa}, {Karim}, {Le F{\`e}vre}, {Lee},
  {Lilly}, {McCracken}, {Michel-Dansac}, {Moutard}, {Renzini}, {Salvato},
  {Sanders}, {Scoville}, {Sheth}, {Silverman}, {Smol{\v{c}}i{\'c}},
  {Taniguchi}, \& {Tresse}}]{Ilbert15}
{Ilbert}, O., {Arnouts}, S., {Le Floc'h}, E., {et~al.} 2015,
  \href{http://dx.doi.org/10.1051/0004-6361/201425176}{\color{magenta}\aap},
  \href{https://ui.adsabs.harvard.edu/abs/2015A&A...579A...2I}{579, A2}

\bibitem[{{Ilbert} {et~al.}(2006){Ilbert}, {Arnouts}, {McCracken},
  {Bolzonella}, {Bertin}, {Le F{\`e}vre}, {Mellier}, {Zamorani}, {Pell{\`o}},
  {Iovino}, {Tresse}, {Le Brun}, {Bottini}, {Garilli}, {Maccagni}, {Picat},
  {Scaramella}, {Scodeggio}, {Vettolani}, {Zanichelli}, {Adami}, {Bardelli},
  {Cappi}, {Charlot}, {Ciliegi}, {Contini}, {Cucciati}, {Foucaud}, {Franzetti},
  {Gavignaud}, {Guzzo}, {Marano}, {Marinoni}, {Mazure}, {Meneux}, {Merighi},
  {Paltani}, {Pollo}, {Pozzetti}, {Radovich}, {Zucca}, {Bondi}, {Bongiorno},
  {Busarello}, {de La Torre}, {Gregorini}, {Lamareille}, {Mathez}, {Merluzzi},
  {Ripepi}, {Rizzo}, \& {Vergani}}]{Ilbert06}
{Ilbert}, O., {Arnouts}, S., {McCracken}, H.~J., {et~al.} 2006,
  \href{http://dx.doi.org/10.1051/0004-6361:20065138}{\color{magenta}\aap},
  \href{https://ui.adsabs.harvard.edu/abs/2006A&A...457..841I}{457, 841}

\bibitem[{{Ilbert} {et~al.}(2010){Ilbert}, {Salvato}, {Le Floc'h}, {Aussel},
  {Capak}, {McCracken}, {Mobasher}, {Kartaltepe}, {Scoville}, {Sanders},
  {Arnouts}, {Bundy}, {Cassata}, {Kneib}, {Koekemoer}, {Le F{\`e}vre}, {Lilly},
  {Surace}, {Taniguchi}, {Tasca}, {Thompson}, {Tresse}, {Zamojski}, {Zamorani},
  \& {Zucca}}]{Ilbert10}
{Ilbert}, O., {Salvato}, M., {Le Floc'h}, E., {et~al.} 2010,
  \href{http://dx.doi.org/10.1088/0004-637X/709/2/644}{\color{magenta}\apj},
  \href{https://ui.adsabs.harvard.edu/abs/2010ApJ...709..644I}{709, 644}

\bibitem[{{Janssen} {et~al.}(2012){Janssen}, {R{\"o}ttgering}, {Best}, \&
  {Brinchmann}}]{Janssen12}
{Janssen}, R.~M.~J., {R{\"o}ttgering}, H.~J.~A., {Best}, P.~N., \&
  {Brinchmann}, J. 2012,
  \href{http://dx.doi.org/10.1051/0004-6361/201219052}{\color{magenta}\aap},
  \href{https://ui.adsabs.harvard.edu/abs/2012A&A...541A..62J}{541, A62}

\bibitem[{{Jarvis} {et~al.}(2015){Jarvis}, {Seymour}, {Afonso}, {Best},
  {Beswick}, {Heywood}, {Huynh}, {Murphy}, {Prandoni}, {Schinnerer}, {Simpson},
  {Vaccari}, \& {White}}]{Jarvis15}
{Jarvis}, M., {Seymour}, N., {Afonso}, J., {et~al.} 2015, in Advancing
  Astrophysics with the Square Kilometre Array (AASKA14),
  \href{https://ui.adsabs.harvard.edu/abs/2015aska.confE..68J}{68}

\bibitem[{{Jim{\'e}nez-Andrade} {et~al.}(2019){Jim{\'e}nez-Andrade},
  {Magnelli}, {Karim}, {Zamorani}, {Bondi}, {Schinnerer}, {Sargent},
  {Romano-D{\'\i}az}, {Novak}, {Lang}, {Bertoldi}, {Vardoulaki}, {Toft},
  {Smol{\v{c}}i{\'c}}, {Harrington}, {Leslie}, {Delhaize}, {Liu}, {Karoumpis},
  {Kartaltepe}, \& {Koekemoer}}]{Jimenez19}
{Jim{\'e}nez-Andrade}, E.~F., {Magnelli}, B., {Karim}, A., {et~al.} 2019,
  \href{http://dx.doi.org/10.1051/0004-6361/201935178}{\color{magenta}\aap},
  \href{https://ui.adsabs.harvard.edu/abs/2019A&A...625A.114J}{625, A114}

\bibitem[{{Jin} {et~al.}(2018){Jin}, {Daddi}, {Liu}, {Smol{\v{c}}i{\'c}},
  {Schinnerer}, {Calabr{\`o}}, {Gu}, {Delhaize}, {Delvecchio}, {Gao},
  {Salvato}, {Puglisi}, {Dickinson}, {Bertoldi}, {Sargent}, {Novak}, {Magdis},
  {Aretxaga}, {Wilson}, \& {Capak}}]{Jin18}
{Jin}, S., {Daddi}, E., {Liu}, D., {et~al.} 2018,
  \href{http://dx.doi.org/10.3847/1538-4357/aad4af}{\color{magenta}\apj},
  \href{https://ui.adsabs.harvard.edu/abs/2018ApJ...864...56J}{864, 56}

\bibitem[{{Karim} {et~al.}(2011){Karim}, {Schinnerer},
  {Mart{\'{\i}}nez-Sansigre}, {Sargent}, {van der Wel}, {Rix}, {Ilbert},
  {Smol{\v c}i{\'c}}, {Carilli}, {Pannella}, {Koekemoer}, {Bell}, \&
  {Salvato}}]{karim11}
{Karim}, A., {Schinnerer}, E., {Mart{\'{\i}}nez-Sansigre}, A., {et~al.} 2011,
  \href{http://dx.doi.org/10.1088/0004-637X/730/2/61}{\color{magenta}\apj},
  \href{https://ui.adsabs.harvard.edu/abs/2011ApJ...730...61K}{730, 61}

\bibitem[{{Kennicutt}(1998)}]{Kennicutt98}
{Kennicutt}, Robert~C., J. 1998,
  \href{http://dx.doi.org/10.1146/annurev.astro.36.1.189}{\color{magenta}\araa},
  \href{https://ui.adsabs.harvard.edu/abs/1998ARA&A..36..189K}{36, 189}

\bibitem[{{Kennicutt} \& {Evans}(2012)}]{kennicutt12}
{Kennicutt}, R.~C. \& {Evans}, N.~J. 2012,
  \href{http://dx.doi.org/10.1146/annurev-astro-081811-125610}{\color{magenta}\araa},
  \href{https://ui.adsabs.harvard.edu/abs/2012ARA%26A..50..531K}{50, 531}

\bibitem[{{Laigle} {et~al.}(2016){Laigle}, {McCracken}, {Ilbert}, {Hsieh},
  {Davidzon}, {Capak}, {Hasinger}, {Silverman}, {Pichon}, {Coupon}, {Aussel},
  {Le Borgne}, {Caputi}, {Cassata}, {Chang}, {Civano}, {Dunlop}, {Fynbo},
  {Kartaltepe}, {Koekemoer}, {Le F{\`e}vre}, {Le Floc'h}, {Leauthaud}, {Lilly},
  {Lin}, {Marchesi}, {Milvang-Jensen}, {Salvato}, {Sanders}, {Scoville},
  {Smolcic}, {Stockmann}, {Taniguchi}, {Tasca}, {Toft}, {Vaccari}, \&
  {Zabl}}]{Laigle16}
{Laigle}, C., {McCracken}, H.~J., {Ilbert}, O., {et~al.} 2016,
  \href{http://dx.doi.org/10.3847/0067-0049/224/2/24}{\color{magenta}\apjs},
  \href{https://ui.adsabs.harvard.edu/abs/2016ApJS..224...24L}{224, 24}

\bibitem[{{Lang} {et~al.}(2016){Lang}, {Hogg}, \& {Mykytyn}}]{Lang16}
{Lang}, D., {Hogg}, D.~W., \& {Mykytyn}, D. 2016, {The Tractor: Probabilistic
  astronomical source detection and measurement}, Astrophysics Source Code
  Library, record ascl:1604.008

\bibitem[{{Le Floc'h} {et~al.}(2009){Le Floc'h}, {Aussel}, {Ilbert},
  {Riguccini}, {Frayer}, {Salvato}, {Arnouts}, {Surace}, {Feruglio},
  {Rodighiero}, {Capak}, {Kartaltepe}, {Heinis}, {Sheth}, {Yan}, {McCracken},
  {Thompson}, {Sanders}, {Scoville}, \& {Koekemoer}}]{lefloch09}
{Le Floc'h}, E., {Aussel}, H., {Ilbert}, O., {et~al.} 2009,
  \href{http://dx.doi.org/10.1088/0004-637X/703/1/222}{\color{magenta}\apj},
  \href{https://ui.adsabs.harvard.edu/abs/2009ApJ...703..222L}{703, 222}

\bibitem[{{Le Floc'h} {et~al.}(2005){Le Floc'h}, {Papovich}, {Dole}, {Bell},
  {Lagache}, {Rieke}, {Egami}, {P{\'e}rez-Gonz{\'a}lez}, {Alonso-Herrero},
  {Rieke}, {Blaylock}, {Engelbracht}, {Gordon}, {Hines}, {Misselt}, {Morrison},
  \& {Mould}}]{LeFloch05}
{Le Floc'h}, E., {Papovich}, C., {Dole}, H., {et~al.} 2005,
  \href{http://dx.doi.org/10.1086/432789}{\color{magenta}\apj},
  \href{https://ui.adsabs.harvard.edu/abs/2005ApJ...632..169L}{632, 169}

\bibitem[{{Leja} {et~al.}(2019){Leja}, {Johnson}, {Conroy}, {van Dokkum},
  {Speagle}, {Brammer}, {Momcheva}, {Skelton}, {Whitaker}, {Franx}, \&
  {Nelson}}]{Leja19}
{Leja}, J., {Johnson}, B.~D., {Conroy}, C., {et~al.} 2019,
  \href{http://dx.doi.org/10.3847/1538-4357/ab1d5a}{\color{magenta}\apj},
  \href{https://ui.adsabs.harvard.edu/abs/2019ApJ...877..140L}{877, 140}

\bibitem[{{Leja} {et~al.}(2017){Leja}, {Johnson}, {Conroy}, {van Dokkum}, \&
  {Byler}}]{leja17}
{Leja}, J., {Johnson}, B.~D., {Conroy}, C., {van Dokkum}, P.~G., \& {Byler}, N.
  2017, \href{http://dx.doi.org/10.3847/1538-4357/aa5ffe}{\color{magenta}\apj},
  \href{https://ui.adsabs.harvard.edu/abs/2017ApJ...837..170L}{837, 170}

\bibitem[{{Leslie} {et~al.}(2020){Leslie}, {Schinnerer}, {Liu}, {Magnelli},
  {Algera}, {Karim}, {Davidzon}, {Gozaliasl}, {Jim{\'e}nez-Andrade}, {Lang},
  {Sargent}, {Novak}, {Groves}, {Smol{\v{c}}i{\'c}}, {Zamorani}, {Vaccari},
  {Battisti}, {Vardoulaki}, {Peng}, \& {Kartaltepe}}]{leslie20}
{Leslie}, S.~K., {Schinnerer}, E., {Liu}, D., {et~al.} 2020,
  \href{http://dx.doi.org/10.3847/1538-4357/aba044}{\color{magenta}\apj},
  \href{https://ui.adsabs.harvard.edu/abs/2020ApJ...899...58L}{899, 58}

\bibitem[{{Lilly} {et~al.}(2013){Lilly}, {Carollo}, {Pipino}, {Renzini}, \&
  {Peng}}]{Lilly13}
{Lilly}, S.~J., {Carollo}, C.~M., {Pipino}, A., {Renzini}, A., \& {Peng}, Y.
  2013,
  \href{http://dx.doi.org/10.1088/0004-637X/772/2/119}{\color{magenta}\apj},
  \href{https://ui.adsabs.harvard.edu/abs/2013ApJ...772..119L}{772, 119}

\bibitem[{{Lisenfeld} \& {V{\"o}lk}(2010)}]{Lisenfeld10}
{Lisenfeld}, U. \& {V{\"o}lk}, H.~J. 2010,
  \href{http://dx.doi.org/10.1051/0004-6361/201015083}{\color{magenta}\aap},
  \href{https://ui.adsabs.harvard.edu/abs/2010A&A...524A..27L}{524, A27}

\bibitem[{{Lutz} {et~al.}(2011){Lutz}, {Poglitsch}, {Altieri}, {Andreani},
  {Aussel}, {Berta}, {Bongiovanni}, {Brisbin}, {Cava}, {Cepa}, {Cimatti},
  {Daddi}, {Dominguez-Sanchez}, {Elbaz}, {F{\"o}rster Schreiber}, {Genzel},
  {Grazian}, {Gruppioni}, {Harwit}, {Le Floc'h}, {Magdis}, {Magnelli},
  {Maiolino}, {Nordon}, {P{\'e}rez Garc{\'\i}a}, {Popesso}, {Pozzi},
  {Riguccini}, {Rodighiero}, {Saintonge}, {Sanchez Portal}, {Santini}, {Shao},
  {Sturm}, {Tacconi}, {Valtchanov}, {Wetzstein}, \& {Wieprecht}}]{Lutz2011}
{Lutz}, D., {Poglitsch}, A., {Altieri}, B., {et~al.} 2011,
  \href{http://dx.doi.org/10.1051/0004-6361/201117107}{\color{magenta}\aap},
  \href{https://ui.adsabs.harvard.edu/abs/2011A&A...532A..90L}{532, A90}

\bibitem[{{Magliocchetti}(2022)}]{Magliocchetti22}
{Magliocchetti}, M. 2022,
  \href{http://dx.doi.org/10.1007/s00159-022-00142-1}{\color{magenta}\aapr},
  \href{https://ui.adsabs.harvard.edu/abs/2022A&ARv..30....6M}{30, 6}

\bibitem[{{Magnelli} {et~al.}(2015){Magnelli}, {Ivison}, {Lutz}, {Valtchanov},
  {Farrah}, {Berta}, {Bertoldi}, {Bock}, {Cooray}, {Ibar}, {Karim}, {Le
  Floc'h}, {Nordon}, {Oliver}, {Page}, {Popesso}, {Pozzi}, {Rigopoulou},
  {Riguccini}, {Rodighiero}, {Rosario}, {Roseboom}, {Wang}, \&
  {Wuyts}}]{Magnelli15}
{Magnelli}, B., {Ivison}, R.~J., {Lutz}, D., {et~al.} 2015,
  \href{http://dx.doi.org/10.1051/0004-6361/201424937}{\color{magenta}\aap},
  \href{https://ui.adsabs.harvard.edu/abs/2015A&A...573A..45M}{573, A45}

\bibitem[{{Magnelli} {et~al.}(2013){Magnelli}, {Popesso}, {Berta}, {Pozzi},
  {Elbaz}, {Lutz}, {Dickinson}, {Altieri}, {Andreani}, {Aussel},
  {B{\'e}thermin}, {Bongiovanni}, {Cepa}, {Charmandaris}, {Chary}, {Cimatti},
  {Daddi}, {F{\"o}rster Schreiber}, {Genzel}, {Gruppioni}, {Harwit}, {Hwang},
  {Ivison}, {Magdis}, {Maiolino}, {Murphy}, {Nordon}, {Pannella}, {P{\'e}rez
  Garc{\'\i}a}, {Poglitsch}, {Rosario}, {Sanchez-Portal}, {Santini}, {Scott},
  {Sturm}, {Tacconi}, \& {Valtchanov}}]{Magnelli13}
{Magnelli}, B., {Popesso}, P., {Berta}, S., {et~al.} 2013,
  \href{http://dx.doi.org/10.1051/0004-6361/201321371}{\color{magenta}\aap},
  \href{https://ui.adsabs.harvard.edu/abs/2013A&A...553A.132M}{553, A132}

\bibitem[{{Ma{\l}ek} {et~al.}(2018){Ma{\l}ek}, {Buat}, {Roehlly}, {Burgarella},
  {Hurley}, {Shirley}, {Duncan}, {Efstathiou}, {Papadopoulos}, {Vaccari},
  {Farrah}, {Marchetti}, \& {Oliver}}]{Malek18}
{Ma{\l}ek}, K., {Buat}, V., {Roehlly}, Y., {et~al.} 2018,
  \href{http://dx.doi.org/10.1051/0004-6361/201833131}{\color{magenta}\aap},
  \href{https://ui.adsabs.harvard.edu/abs/2018A&A...620A..50M}{620, A50}

\bibitem[{{Mohan} \& {Rafferty}(2015)}]{PyBDSF07}
{Mohan}, N. \& {Rafferty}, D. 2015, {PyBDSF: Python Blob Detection and Source
  Finder}, Astrophysics Source Code Library, record ascl:1502.007

\bibitem[{{Momcheva} {et~al.}(2016){Momcheva}, {Brammer}, {van Dokkum},
  {Skelton}, {Whitaker}, {Nelson}, {Fumagalli}, {Maseda}, {Leja}, {Franx},
  {Rix}, {Bezanson}, {Da Cunha}, {Dickey}, {F{\"o}rster Schreiber},
  {Illingworth}, {Kriek}, {Labb{\'e}}, {Ulf Lange}, {Lundgren}, {Magee},
  {Marchesini}, {Oesch}, {Pacifici}, {Patel}, {Price}, {Tal}, {Wake}, {van der
  Wel}, \& {Wuyts}}]{Momcheva:2016lr}
{Momcheva}, I.~G., {Brammer}, G.~B., {van Dokkum}, P.~G., {et~al.} 2016,
  \href{http://dx.doi.org/10.3847/0067-0049/225/2/27}{\color{magenta}\apjs},
  \href{http://adsabs.harvard.edu/abs/2016ApJS..225...27M}{225, 27}

\bibitem[{{Moutard} {et~al.}(2020){Moutard}, {Sawicki}, {Arnouts}, {Golob},
  {Coupon}, {Ilbert}, {Yang}, \& {Gwyn}}]{Moutard20}
{Moutard}, T., {Sawicki}, M., {Arnouts}, S., {et~al.} 2020,
  \href{http://dx.doi.org/10.1093/mnras/staa706}{\color{magenta}\mnras},
  \href{https://ui.adsabs.harvard.edu/abs/2020MNRAS.494.1894M}{494, 1894}

\bibitem[{{Murphy}(2009)}]{Murphy09}
{Murphy}, E.~J. 2009,
  \href{http://dx.doi.org/10.1088/0004-637X/706/1/482}{\color{magenta}\apj},
  \href{https://ui.adsabs.harvard.edu/abs/2009ApJ...706..482M}{706, 482}

\bibitem[{{Noll} {et~al.}(2009){Noll}, {Burgarella}, {Giovannoli}, {Buat},
  {Marcillac}, \& {Mu{\~n}oz-Mateos}}]{Noll09}
{Noll}, S., {Burgarella}, D., {Giovannoli}, E., {et~al.} 2009,
  \href{http://dx.doi.org/10.1051/0004-6361/200912497}{\color{magenta}\aap},
  \href{https://ui.adsabs.harvard.edu/abs/2009A&A...507.1793N}{507, 1793}

\bibitem[{Oke(1974)}]{oke_absolute_1974}
Oke, J.~B. 1974, \href{http://dx.doi.org/10.1086/190287}{\color{magenta}ApJS},
  27, 21

\bibitem[{{Oliver} {et~al.}(2012){Oliver}, {Bock}, {Altieri}, {Amblard},
  {Arumugam}, {Aussel}, {Babbedge}, {Beelen}, {B{\'e}thermin}, {Blain},
  {Boselli}, {Bridge}, {Brisbin}, {Buat}, {Burgarella},
  {Castro-Rodr{\'\i}guez}, {Cava}, {Chanial}, {Cirasuolo}, {Clements},
  {Conley}, {Conversi}, {Cooray}, {Dowell}, {Dubois}, {Dwek}, {Dye}, {Eales},
  {Elbaz}, {Farrah}, {Feltre}, {Ferrero}, {Fiolet}, {Fox}, {Franceschini},
  {Gear}, {Giovannoli}, {Glenn}, {Gong}, {Gonz{\'a}lez Solares}, {Griffin},
  {Halpern}, {Harwit}, {Hatziminaoglou}, {Heinis}, {Hurley}, {Hwang}, {Hyde},
  {Ibar}, {Ilbert}, {Isaak}, {Ivison}, {Lagache}, {Le Floc'h}, {Levenson},
  {Faro}, {Lu}, {Madden}, {Maffei}, {Magdis}, {Mainetti}, {Marchetti},
  {Marsden}, {Marshall}, {Mortier}, {Nguyen}, {O'Halloran}, {Omont}, {Page},
  {Panuzzo}, {Papageorgiou}, {Patel}, {Pearson}, {P{\'e}rez-Fournon}, {Pohlen},
  {Rawlings}, {Raymond}, {Rigopoulou}, {Riguccini}, {Rizzo}, {Rodighiero},
  {Roseboom}, {Rowan-Robinson}, {S{\'a}nchez Portal}, {Schulz}, {Scott},
  {Seymour}, {Shupe}, {Smith}, {Stevens}, {Symeonidis}, {Trichas}, {Tugwell},
  {Vaccari}, {Valtchanov}, {Vieira}, {Viero}, {Vigroux}, {Wang}, {Ward},
  {Wardlow}, {Wright}, {Xu}, \& {Zemcov}}]{Oliver12}
{Oliver}, S.~J., {Bock}, J., {Altieri}, B., {et~al.} 2012,
  \href{http://dx.doi.org/10.1111/j.1365-2966.2012.20912.x}{\color{magenta}\mnras},
  \href{https://ui.adsabs.harvard.edu/abs/2012MNRAS.424.1614O}{424, 1614}

\bibitem[{{Padovani}(2016)}]{Padovani16}
{Padovani}, P. 2016,
  \href{http://dx.doi.org/10.1007/s00159-016-0098-6}{\color{magenta}\aapr},
  \href{https://ui.adsabs.harvard.edu/abs/2016A&ARv..24...13P}{24, 13}

\bibitem[{{Pracy} {et~al.}(2016){Pracy}, {Ching}, {Sadler}, {Croom}, {Baldry},
  {Bland-Hawthorn}, {Brough}, {Brown}, {Couch}, {Davis}, {Drinkwater},
  {Hopkins}, {Jarvis}, {Jelliffe}, {Jurek}, {Loveday}, {Pimbblet}, {Prescott},
  {Wisnioski}, \& {Woods}}]{Pracy16}
{Pracy}, M.~B., {Ching}, J. H.~Y., {Sadler}, E.~M., {et~al.} 2016,
  \href{http://dx.doi.org/10.1093/mnras/stw910}{\color{magenta}\mnras},
  \href{https://ui.adsabs.harvard.edu/abs/2016MNRAS.460....2P}{460, 2}

\bibitem[{{Rajpurohit} {et~al.}(2022){Rajpurohit}, {Hoeft}, {Wittor}, {van
  Weeren}, {Vazza}, {Rudnick}, {Rajpurohit}, {Forman}, {Riseley}, {Brienza},
  {Bonafede}, {Rajpurohit}, {Dom{\'\i}nguez-Fern{\'a}ndez}, {Eilek},
  {Bonnassieux}, {Br{\"u}ggen}, {Loi}, {R{\"o}ttgering}, {Drabent},
  {Locatelli}, {Botteon}, {Brunetti}, \& {Clarke}}]{Rajpurohit_2021}
{Rajpurohit}, K., {Hoeft}, M., {Wittor}, D., {et~al.} 2022,
  \href{http://dx.doi.org/10.1051/0004-6361/202142340}{\color{magenta}\aap},
  \href{https://ui.adsabs.harvard.edu/abs/2022A&A...657A...2R}{657, A2}

\bibitem[{Reichherzer {et~al.}(2021)Reichherzer, Merten, Dörner, Tjus,
  Pueschel, \& Zweibel}]{Reichherzer_2021}
Reichherzer, P., Merten, L., Dörner, J., {et~al.} 2021,
  \href{http://dx.doi.org/10.1007/s42452-021-04891-z}{\color{magenta}{SN}
  Applied Sciences}, 4

\bibitem[{{Sailer}(2021)}]{Sailor21}
{Sailer}, M.~W. 2021, {A James Webb Space Telescope NIRCam Deep Field
  Simplified Simulation Using a Geometric-Focused Ensemble Approach}

\bibitem[{{Saito} {et~al.}(2020){Saito}, {de la Torre}, {Ilbert}, {Dubois},
  {Yabe}, \& {Coupon}}]{Saito20}
{Saito}, S., {de la Torre}, S., {Ilbert}, O., {et~al.} 2020,
  \href{http://dx.doi.org/10.1093/mnras/staa727}{\color{magenta}\mnras},
  \href{https://ui.adsabs.harvard.edu/abs/2020MNRAS.494..199S}{494, 199}

\bibitem[{{Schleicher} \& {Beck}(2013)}]{Schleicher13}
{Schleicher}, D. R.~G. \& {Beck}, R. 2013,
  \href{http://dx.doi.org/10.1051/0004-6361/201321707}{\color{magenta}\aap},
  \href{https://ui.adsabs.harvard.edu/abs/2013A&A...556A.142S}{556, A142}

\bibitem[{{Schreiber} {et~al.}(2018){Schreiber}, {Labb{\'e}}, {Glazebrook},
  {Bekiaris}, {Papovich}, {Costa}, {Elbaz}, {Kacprzak}, {Nanayakkara}, {Oesch},
  {Pannella}, {Spitler}, {Straatman}, {Tran}, \& {Wang}}]{Schreiber18}
{Schreiber}, C., {Labb{\'e}}, I., {Glazebrook}, K., {et~al.} 2018,
  \href{http://dx.doi.org/10.1051/0004-6361/201731917}{\color{magenta}\aap},
  \href{https://ui.adsabs.harvard.edu/abs/2018A&A...611A..22S}{611, A22}

\bibitem[{{Silverman} {et~al.}(2009){Silverman}, {Lamareille}, {Maier},
  {Lilly}, {Mainieri}, {Brusa}, {Cappelluti}, {Hasinger}, {Zamorani},
  {Scodeggio}, {Bolzonella}, {Contini}, {Carollo}, {Jahnke}, {Kneib}, {Le
  F{\`e}vre}, {Merloni}, {Bardelli}, {Bongiorno}, {Brunner}, {Caputi},
  {Civano}, {Comastri}, {Coppa}, {Cucciati}, {de la Torre}, {de Ravel},
  {Elvis}, {Finoguenov}, {Fiore}, {Franzetti}, {Garilli}, {Gilli}, {Iovino},
  {Kampczyk}, {Knobel}, {Kova{\v c}}, {Le Borgne}, {Le Brun}, {Mignoli},
  {Pello}, {Peng}, {Perez Montero}, {Ricciardelli}, {Tanaka}, {Tasca},
  {Tresse}, {Vergani}, {Vignali}, {Zucca}, {Bottini}, {Cappi}, {Cassata},
  {Fumana}, {Griffiths}, {Kartaltepe}, {Koekemoer}, {Marinoni}, {McCracken},
  {Memeo}, {Meneux}, {Oesch}, {Porciani}, \& {Salvato}}]{Silverman:2009}
{Silverman}, J.~D., {Lamareille}, F., {Maier}, C., {et~al.} 2009,
  \href{http://dx.doi.org/10.1088/0004-637X/696/1/396}{\color{magenta}\apj},
  \href{http://cdsads.u-strasbg.fr/abs/2009ApJ...696..396S}{696, 396}

\bibitem[{{Smol{\v{c}}i{\'c}} {et~al.}(2017{\natexlab{a}}){Smol{\v{c}}i{\'c}},
  {Novak}, {Bondi}, {Ciliegi}, {Mooley}, {Schinnerer}, {Zamorani}, {Navarrete},
  {Bourke}, {Karim}, {Vardoulaki}, {Leslie}, {Delhaize}, {Carilli}, {Myers},
  {Baran}, {Delvecchio}, {Miettinen}, {Banfield}, {Balokovi{\'c}}, {Bertoldi},
  {Capak}, {Frail}, {Hallinan}, {Hao}, {Herrera Ruiz}, {Horesh}, {Ilbert},
  {Intema}, {Jeli{\'c}}, {Kl{\"o}ckner}, {Krpan}, {Kulkarni}, {McCracken},
  {Laigle}, {Middleberg}, {Murphy}, {Sargent}, {Scoville}, \&
  {Sheth}}]{Smolcic17}
{Smol{\v{c}}i{\'c}}, V., {Novak}, M., {Bondi}, M., {et~al.} 2017{\natexlab{a}},
  \href{http://dx.doi.org/10.1051/0004-6361/201628704}{\color{magenta}\aap},
  \href{https://ui.adsabs.harvard.edu/abs/2017A&A...602A...1S}{602, A1}

\bibitem[{{Smol{\v{c}}i{\'c}} {et~al.}(2017{\natexlab{b}}){Smol{\v{c}}i{\'c}},
  {Novak}, {Bondi}, {Ciliegi}, {Mooley}, {Schinnerer}, {Zamorani}, {Navarrete},
  {Bourke}, {Karim}, {Vardoulaki}, {Leslie}, {Delhaize}, {Carilli}, {Myers},
  {Baran}, {Delvecchio}, {Miettinen}, {Banfield}, {Balokovi{\'c}}, {Bertoldi},
  {Capak}, {Frail}, {Hallinan}, {Hao}, {Herrera Ruiz}, {Horesh}, {Ilbert},
  {Intema}, {Jeli{\'c}}, {Kl{\"o}ckner}, {Krpan}, {Kulkarni}, {McCracken},
  {Laigle}, {Middleberg}, {Murphy}, {Sargent}, {Scoville}, \&
  {Sheth}}]{smolcic17b}
{Smol{\v{c}}i{\'c}}, V., {Novak}, M., {Bondi}, M., {et~al.} 2017{\natexlab{b}},
  \href{http://dx.doi.org/10.1051/0004-6361/201628704}{\color{magenta}\aap},
  \href{https://ui.adsabs.harvard.edu/abs/2017A&A...602A...1S}{602, A1}

\bibitem[{{Smol{\v{c}}i{\'c}} {et~al.}(2017{\natexlab{c}}){Smol{\v{c}}i{\'c}},
  {Novak}, {Delvecchio}, {Ceraj}, {Bondi}, {Delhaize}, {Marchesi}, {Murphy},
  {Schinnerer}, {Vardoulaki}, \& {Zamorani}}]{smolcic17a}
{Smol{\v{c}}i{\'c}}, V., {Novak}, M., {Delvecchio}, I., {et~al.}
  2017{\natexlab{c}},
  \href{http://dx.doi.org/10.1051/0004-6361/201730685}{\color{magenta}\aap},
  \href{https://ui.adsabs.harvard.edu/abs/2017A&A...602A...6S}{602, A6}

\bibitem[{{Smol{\v{c}}i{\'c}} {et~al.}(2008){Smol{\v{c}}i{\'c}}, {Schinnerer},
  {Scodeggio}, {Franzetti}, {Aussel}, {Bondi}, {Brusa}, {Carilli}, {Capak},
  {Charlot}, {Ciliegi}, {Ilbert}, {Ivezi{\'c}}, {Jahnke}, {McCracken},
  {Obri{\'c}}, {Salvato}, {Sanders}, {Scoville}, {Trump}, {Tremonti}, {Tasca},
  {Walcher}, \& {Zamorani}}]{Smolcic08}
{Smol{\v{c}}i{\'c}}, V., {Schinnerer}, E., {Scodeggio}, M., {et~al.} 2008,
  \href{http://dx.doi.org/10.1086/588028}{\color{magenta}\apjs},
  \href{https://ui.adsabs.harvard.edu/abs/2008ApJS..177...14S}{177, 14}

\bibitem[{{Snyder} {et~al.}(2019){Snyder}, {Rodriguez-Gomez}, {Lotz}, {Torrey},
  {Quirk}, {Hernquist}, {Vogelsberger}, \& {Freeman}}]{Snyder_2019}
{Snyder}, G.~F., {Rodriguez-Gomez}, V., {Lotz}, J.~M., {et~al.} 2019,
  \href{http://dx.doi.org/10.1093/mnras/stz1059}{\color{magenta}\mnras},
  \href{https://ui.adsabs.harvard.edu/abs/2019MNRAS.486.3702S}{486, 3702}

\bibitem[{{Stalevski} {et~al.}(2016){Stalevski}, {Ricci}, {Ueda}, {Lira},
  {Fritz}, \& {Baes}}]{Stalevski16}
{Stalevski}, M., {Ricci}, C., {Ueda}, Y., {et~al.} 2016,
  \href{http://dx.doi.org/10.1093/mnras/stw444}{\color{magenta}\mnras},
  \href{https://ui.adsabs.harvard.edu/abs/2016MNRAS.458.2288S}{458, 2288}

\bibitem[{{Szokoly} {et~al.}(2004){Szokoly}, {Bergeron}, {Hasinger}, {Lehmann},
  {Kewley}, {Mainieri}, {Nonino}, {Rosati}, {Giacconi}, {Gilli}, {Gilmozzi},
  {Norman}, {Romaniello}, {Schreier}, {Tozzi}, {Wang}, {Zheng}, \&
  {Zirm}}]{Szokoly04}
{Szokoly}, G.~P., {Bergeron}, J., {Hasinger}, G., {et~al.} 2004,
  \href{http://dx.doi.org/10.1086/424707}{\color{magenta}\apjs},
  \href{https://ui.adsabs.harvard.edu/abs/2004ApJS..155..271S}{155, 271}

\bibitem[{{Tabatabaei} {et~al.}(2017){Tabatabaei}, {Schinnerer}, {Krause},
  {Dumas}, {Meidt}, {Damas-Segovia}, {Beck}, {Murphy}, {Mulcahy}, {Groves},
  {Bolatto}, {Dale}, {Galametz}, {Sandstrom}, {Boquien}, {Calzetti},
  {Kennicutt}, {Hunt}, {De Looze}, \& {Pellegrini}}]{Tabatabaei17}
{Tabatabaei}, F.~S., {Schinnerer}, E., {Krause}, M., {et~al.} 2017,
  \href{http://dx.doi.org/10.3847/1538-4357/836/2/185}{\color{magenta}\apj},
  \href{https://ui.adsabs.harvard.edu/abs/2017ApJ...836..185T}{836, 185}

\bibitem[{{Tacchella} {et~al.}(2022){Tacchella}, {Finkelstein}, {Bagley},
  {Dickinson}, {Ferguson}, {Giavalisco}, {Graziani}, {Grogin}, {Hathi},
  {Hutchison}, {Jung}, {Koekemoer}, {Larson}, {Papovich}, {Pirzkal},
  {Rojas-Ruiz}, {Song}, {Schneider}, {Somerville}, {Wilkins}, \&
  {Yung}}]{Tacchella22}
{Tacchella}, S., {Finkelstein}, S.~L., {Bagley}, M., {et~al.} 2022,
  \href{http://dx.doi.org/10.3847/1538-4357/ac4cad}{\color{magenta}\apj},
  \href{https://ui.adsabs.harvard.edu/abs/2022ApJ...927..170T}{927, 170}

\bibitem[{{Walcher} {et~al.}(2011){Walcher}, {Groves}, {Budav{\'a}ri}, \&
  {Dale}}]{Walcher11}
{Walcher}, J., {Groves}, B., {Budav{\'a}ri}, T., \& {Dale}, D. 2011,
  \href{http://dx.doi.org/10.1007/s10509-010-0458-z}{\color{magenta}\apss},
  \href{https://ui.adsabs.harvard.edu/abs/2011Ap&SS.331....1W}{331, 1}

\bibitem[{{Weaver} {et~al.}(2022){Weaver}, {Kauffmann}, {Ilbert}, {McCracken},
  {Moneti}, {Toft}, {Brammer}, {Shuntov}, {Davidzon}, {Hsieh}, {Laigle},
  {Anastasiou}, {Jespersen}, {Vinther}, {Capak}, {Casey}, {McPartland},
  {Milvang-Jensen}, {Mobasher}, {Sanders}, {Zalesky}, {Arnouts}, {Aussel},
  {Dunlop}, {Faisst}, {Franx}, {Furtak}, {Fynbo}, {Gould}, {Greve}, {Gwyn},
  {Kartaltepe}, {Kashino}, {Koekemoer}, {Kokorev}, {Le F{\`e}vre}, {Lilly},
  {Masters}, {Magdis}, {Mehta}, {Peng}, {Riechers}, {Salvato}, {Sawicki},
  {Scarlata}, {Scoville}, {Shirley}, {Silverman}, {Sneppen}, {Smolc̆i{\'c}},
  {Steinhardt}, {Stern}, {Tanaka}, {Taniguchi}, {Teplitz}, {Vaccari}, {Wang},
  \& {Zamorani}}]{Weaver_2022}
{Weaver}, J.~R., {Kauffmann}, O.~B., {Ilbert}, O., {et~al.} 2022,
  \href{http://dx.doi.org/10.3847/1538-4365/ac3078}{\color{magenta}\apjs},
  \href{https://ui.adsabs.harvard.edu/abs/2022ApJS..258...11W}{258, 11}

\bibitem[{{Williams} \& {R{\"o}ttgering}(2015)}]{Williams15}
{Williams}, W.~L. \& {R{\"o}ttgering}, H.~J.~A. 2015,
  \href{http://dx.doi.org/10.1093/mnras/stv692}{\color{magenta}\mnras},
  \href{https://ui.adsabs.harvard.edu/abs/2015MNRAS.450.1538W}{450, 1538}

\bibitem[{{Yang} {et~al.}(2020){Yang}, {Boquien}, {Buat}, {Burgarella},
  {Ciesla}, {Duras}, {Stalevski}, {Brandt}, \& {Papovich}}]{Yang19}
{Yang}, G., {Boquien}, M., {Buat}, V., {et~al.} 2020,
  \href{http://dx.doi.org/10.1093/mnras/stz3001}{\color{magenta}\mnras},
  \href{https://ui.adsabs.harvard.edu/abs/2020MNRAS.491..740Y}{491, 740}

\end{thebibliography}
\onecolumn
\end{document}